\documentclass[epjCONF]{svjour}
\usepackage{graphicx,subfigure}
\usepackage[varg]{txfonts} % Times fonts
\usepackage[latin1]{inputenc}
\newcommand{\beq}{\begin{equation}}
\newcommand{\eeq}{\end{equation}}
\newcommand{\ba}{\begin{eqnarray}}
\newcommand{\ea}{\end{eqnarray}}

\def\lsim{\raise0.3ex\hbox{$\;<$\kern-0.75em\raise-1.1ex\hbox{$\sim\;$}}}
\def\gsim{\raise0.3ex\hbox{$\;>$\kern-0.75em\raise-1.1ex\hbox{$\sim\;$}}}

\def\theta{\vartheta}

\session-title{Future Direcrions in UHECR Physics 2012}
\begin{document}
\title{UHECR: Signatures and Models}
\author{V. Berezinsky}
\institute{INFN, National Gran Sasso Laboratory, I-67010, Assergi
(AQ) Italy}
\abstract{ The signatures of Ultra High Energy ($E \gtrsim 1$~ EeV) proton
propagation through CMB radiation are pair-production dip and GZK
cutoff.  The visible characteristics of these two spectral features
are ankle, which is intrinsic part of the dip, beginning of GZK cutoff
in the differential spectrum and $E_{1/2}$ in integral spectrum.
Measured by HiRes and Telescope Array (TA) these characteristics agree
with theoretical predictions. However, directly measured mass
composition remains a puzzle. While HiRes and TA detectors observe the
proton-dominated mass composition, the data of Auger detector
strongly evidence for nuclei mass composition becoming progressively
heavier at energy higher than 4~EeV and reaching Iron at energy about
35~EeV. The models based on the Auger and HiRes/TA data are considered
independently and classified using the transition from galactic to
extragalactic cosmic rays. The ankle cannot provide this transition.
since data of all three detectors at energy (1 - 3)~EeV agree with
pure proton composition (or at least not heavier than Helium). If
produced in Galaxy these particles result in too high anisotropy.
This argument excludes or strongly disfavours all ankle models with
ankle energy $E_a > 3$~EeV. The calculation of elongation curves,
$X_{\max}(E)$, for different ankle models strengthens further this
conclusion. Status of other models, the dip, mixed composition and
Auger based models are discussed.
}
\maketitle
\section{Introduction}
\label{introduction}
The observed energy spectrum of Cosmic Rays (CR) has an approximately
power-law behavior for $11$ orders of magnitude in energy with several
features that can be linked with particles propagation and acceleration.
This power-law behavior is most probably indicative of a power-law
acceleration spectra, while spectral features may be assigned to
changes in the origin of particles, their propagation and acceleration.

The most prominent feature is the {\em knee} in all-particle spectrum
at energy 3-4~PeV, discovered first at the MSU (Moscow State University)
array in 1958 \cite{MSU-knee}. At the knee the spectrum $E^{-\gamma}$
steepens from $\gamma \approx 2.7$ to $\gamma \approx 3.1$. This knee
is provided by the light elements, protons and Helium, and is
explained in the framework of the Standard Model (SM) for Galactic Cosmic
Rays (GCR) by the maximum energy $E_{\max}$ of acceleration in the Galactic
Sources  (Supernovae Remnants, SNRs). In the case of the rigidity-dependent
acceleration  $E_{\max} \propto Z$, where $Z$ is charge number of a
nuclei, the maximum acceleration energy  is reached by Iron nuclei and
the Iron knee is predicted to be located at energy by factor 26 times
higher than for proton knee, i.e. at energy $E_{\max}^{\rm Fe} \sim
0.1$~EeV. Recently, the Iron knee was found indeed at energy 0.8~EeV
in KASCADE-Grande experiment \cite{kascadeFe} in a good agreement
with rigidity acceleration prediction.

Above the knee, at energy $E_{\rm skn} \approx (0.4 - 0.7)$~EeV, there
is a faint feature in the spectrum \cite{second-knee} called the {\em
second knee}. It is seen in many experiments (for a review see
\cite{2knee-ankle-rev}). This feature is often interpreted as the
place of transition from galactic to extragalactic CRs.

However, for the last forty years the standard place for transition
from galactic to extragalactic CRs is considered at {\em ankle},
a very prominent spectral feature, observed first
in 1960s at Volcano Ranch detector, and  immediately
interpreted by J. Linsley \cite{linsley} as transition between
these two components of CRs. At present beginning of ankle is
determined as $E_a^{\rm obs} \approx 4.6$~ EeV according to HiRes
observations \cite{ankle-hires}, and $E_a^{\rm obs}
\approx (4.0 - 4.5)$~ EeV according to Auger (PAO) observations
\cite{ankle-PAO,Kampert}.

The highest energy feature, the steepening ('cutoff') of the
spectrum is found  in all three largest detectors, HiRes
\cite{GZK-hires}, Telescope Array (TA) \cite{GZK-TA} and
PAO \cite{GZK-PAO}, though the nature of this cutoff is still
questionable.

The spectral features expected theoretically are much different for
UHE protons and nuclei.

If primary particles are protons, their spectrum must show a steepening
which begins at $E_{\rm GZK} \sim 50$~EeV. This is the famous
Greisen \cite{greisen}, Zatsepin and Kuzmin \cite{ZK} cutoff,
referred to as GZK, caused by pion photo-production energy loss
in collisions of protons with the CMB photons:
\beq
p + \gamma_{\rm cmb} \rightarrow N + \pi
\label{eq:gzk}
\eeq
Most probably this cutoff is already
observed in the HiRes \cite{GZK-hires} and TA \cite{GZK-TA} data,
while the spectrum steepening observed by the PAO \cite{GZK-PAO}
does not agree well, according to our calculations, with the predicted
GZK shape and position.

There is another feature in the spectrum of UHE extragalactic protons,
the {\em pair-production dip}, which, as in case of GZK cutoff,
can be directly linked to the interaction of CRs with the CMB. This dip
arises due to electron-positron production energy loss by extragalactic
protons interacting with the CMB photons:
\begin{equation}
p+\gamma_{\rm CMB} \rightarrow p+e^++e^- .
\label{eq:pairprod}
\end{equation}

This feature has been studied in \cite{BG88,Stanev2000,BGGPL,BGGprd,Aletal}.
The dip was observed with a very good statistical significance,
$\chi^2/\mbox{d.o.f.} \sim 1$, by the Fly's Eye, Yakutsk, Akeno-AGASA
and HiRes detectors, but it is absent in the Auger (PAO)
data (see section \ref{sec:auger} for a more detailed analysis).
A remarkable property of the pair-production dip is an automatic
description of the ankle, which appears as a flat intrinsic part of the dip.

The pair-production dip and GZK cutoff are signatures of protons. A
confirmation of the shape of these features can be considered as an
indirect evidence for a proton-dominated composition of primary CRs.
For nuclei as primaries the shapes of the dip and cutoff are much
different.

A different explanation of the dip has been proposed by Hill and Schramm
\cite{HS85}. They interpreted the dip observed in 1980s in terms of a
two-component model; the low energy component was either galactic or
produced by Local Supercluster. A similar model was later considered in
\cite{YT}. The Hill-Schramm's interpretation is widely accepted now.

An alternative to the protons as primaries is given by nuclei.
Propagating through astrophysical backgrounds, nuclei lose
their energy in photo-disintegration and pair production processes, and
also due to adiabatic expansion of the universe. The steepening of nuclei
spectra due to photo-disintegration in the interactions with the CMB has
been shortly mentioned by Greisen \cite{greisen} and later considered
quantitatively in \cite{stecker69,BZ71,BGZ75,hillas75,stecker75}. The
beginning of the steepening in this case (nuclei in the CMB radiation only)
corresponds to the intersection of $e^-e^+$ and adiabatic energy losses
 curves \cite{BZ71,BGZ75}. The beginning of this steepening is lower than in
the case of GZK cutoff and the shape of this feature differs from
that of GZK. Nowadays the importance of Extragalactic Background
Light (EBL) becomes evident. Interaction with high energy EBL
photons results in photo-disintegration of nuclei at low energies,
while in pair-production energy losses and photo-disintegration at
highest energies, CMB radiation dominates. Moreover, as recent
accurate calculations show the PAO spectrum steepening at highest
energies may be explained not only by nuclei interactions. It may
be also produced as a result of a decreasing acceleration
efficiency in the vicinity of the maximum energy $E_{\max}$ that 
sources can provide  (for reviews see \cite{GZK-rev}). The HiRes
and TA data, in favour of proton composition, and the PAO data, in
favour of nuclei composition, are further discussed in
sections~\ref{sec:GZK}  and~\ref{sec:auger}.

The GZK cutoff and pair-production dip are signatures of protons as
primary particles. However. the direct measurement of mass composition
is needed for  consistency  of the whole picture. The most reliable
method to measure the mass composition nowadays is given by determination
of the depth in the atmosphere $X_{\max}$ where the extensive air
shower (EAS) reaches maximum and the distribution of $X_{\max}$
over average value given by RMS($X_{\max}$). The data of
HiRes \cite{ankle-hires} and TA \cite{TAicrc11}
show the proton-dominant mass composition at energy $E \gsim 1$~EeV
in accordance with pair-production dip and GZK cutoff, while PAO
data show nuclei mass composition with steadily increasing mass number
$A$ at $E\gsim 3$~EeV, which practically reaches Iron at $E \sim 35$~EeV.
This discrepancy of observational data makes us to study two different
scenarios with (almost) pure proton composition, based on HiRes and TA
data, and Auger-based scenario with steadily heavier mass composition
at increasing energy.  Four phenomenological models will be considered.
The distinctive feature which characterizes each of them is energy
of transition from Galactic to extragalactic cosmic rays. The main
prediction is the mass composition and energy spectrum.

In the {\em ankle models} \cite{stanev2005} - \cite{waxman} it is assumed
that the transition occurs at the flat part of the observed spectrum
in the energy interval $E_a^{\rm trans} \sim (3 - 10)$~EeV. The
transition energy  is given by the
intersection of a flat extragalactic spectrum and a very steep galactic
one. In the majority of ankle models the extragalactic component is
assumed to be pure proton, while the galactic one should be naturally
represented by Iron nuclei at energies above the Iron knee. These
models predict a transition from an Iron-dominated composition to a
proton-dominated one at the ankle energy.

In the {\em mixed composition model} \cite{mixed} it is assumed that
the extragalactic component consists of nuclei of various types. Thus
transition here occurs from Iron to lighter nuclei of mixed composition;
it can occur at the ankle or nearby it.

In the {\em dip model} the transition begins at the second knee and is
completed at the beginning of the dip, at $E \approx 1 $~EeV. The ankle
in this model appears as an intrinsic part of the dip. Like in the ankle
model, the transition here also occurs as an intersection of the flat
extragalactic component (this flatness is especially prominent in the
case of diffusive propagation) with the steep Galactic spectrum. In
contrast to the ankle and mixed composition models, the dip model
predicts an almost pure proton composition above $E \approx 1$~EeV and a
pure Iron composition below this energy.

The {\em Auger-based models}  are build for explanation of PAO data:
energy spectrum \cite{auger-spectrum} and mass
composition \cite{auger-mass} with proton or Helium dominance at
$(1 - 3)$~EeV and with nuclei at higher energies with steadily
increasing atomic number $A$ with energy. Transition from Galactic to
extragalactic CRs occurs at the ankle.

We discuss in this presentation the status of these models.

\section{Signatures of proton propagation through CMB}
\label{signatures}

Propagating through CMB, UHE protons undergo two interactions:
photo-pion production $p+\gamma_{\rm cmb}\to N+ \pi$ and
$e^+e^-$ production $p+\gamma_{\rm cmb}\to p+e^++e^-$.  As a result,
the proton spectrum is distorted: due to the first interaction
it obtains a sharp steepening called {\em GZK cutoff} and due to the
second one - shallow deepening, called {\em dip}. Both features depend
not only on interactions but also on model-dependent quantities, e.g.
on modes of propagation (diffusion or rectilinear propagation), on
cosmological evolution of the sources, on source separation etc. These
model-dependent distortions are especially strong for GZK feature.

{\em The main strategy of this presentation as well as of works
\cite{BGGPL,BGGprd,Aletal} is to distinguish the interaction signatures
from the model-dependent ones.}

This is possible to do using the {\em modification factor}
$\eta(E)$ \cite{BG88}, some kind
of theoretical spectrum, in which model-dependent features are suppressed
or absent. For this aim we use the calculations which involves only one
free parameter. One naturally expects that the calculated interaction
signature in terms of modification factor cannot have the agreement with
observations with good $\chi^2$, because the observational data include
the model-dependent features described by many parameters, such as
cosmological evolution of the sources, source separation etc. One free
parameter is not enough to describe 4 - 5 different experiments with
about 20 energy bins in each. As the next step we perform the
{\em model-dependent} calculations which necessarily include many free
parameters improving further the agreement. This analysis should
be done in terms of usual $E^3J(E)$ spectrum, where the model-dependent
features are not suppressed.

However, the Nature has been more kind to us than we expected. The
dip, in terms of modification factor with one free physical parameter
(interaction-signature description) gave very good $\chi^2$ agreement
with observations of four experiments \cite{data}: Yakutsk, AGASA,
HiRes and Telescope Array (see section \ref{sec:modfactor-dip}).
The comparison with PAO data has a different story. Comparison in
terms of modification factor with PAO observational data from ICRC
2007 (Merida) had good enough $\chi^2$ (see \cite{TAUP2007}). However,
with increasing statistics the agreement became worse, and comparison
of modification factor with PAO data 2010 - 2001 has bad $\chi^2$. It
could be that for the difference in very small error bars the
model-dependent effects are responsible. Indeed, as demonstrated in
section \ref{sec:energy-calibrator} the 22\% shift of PAO energy scale
and cosmological
evolution make the PAO spectrum in terms of  $E^3J(E)$  compatible with
the pair-production dip with high accuracy. However, this interpretation
contradicts the nuclei mass composition measured by PAO in energy region
of the dip.

\subsection{The dip in terms of modification factor}
\label{sec:modfactor-dip} In this section the dip will be studied
as a {\em signature} of UHE proton interaction with CMB, using the
modification factor as a tool. The modification factor $\eta(E)$
is defined as the ratio of proton spectrum $J_p(E)$ calculated
with all energy losses to the so-called unmodified spectrum
$J_{\rm unm}(E)$ in which only adiabatic energy losses (red-shift)
are included: \beq \eta(E)=J_p(E)/J_{\rm unm}(E) \label{eq:unm}
\eeq Modification factor is an excellent characteristic of {\em
interaction signature}. As one might see the interactions enter
only numerator and thus they are not suppressed in $\eta(E)$,
while most other phenomena enter both numerator and denominator
and thus they are suppressed or even cancelled in modification
factor. property is especially pronounced for the dip
modification factor, which according to our calculations
\cite{Aletal}depends very weakly on generation index $\gamma_g$
and $E_{\max}$, on propagation mode, source separation within 1 -
50 Mpc, local source overdensity or deficit etc.  The dip
modification factor is modified strongly by presence of nuclei
($\gsim 15\%$). Modification factor for GZK feature is changing
stronger.
\begin{figure}
\begin{center}
\resizebox{0.5\textwidth}{!}{%
 \includegraphics{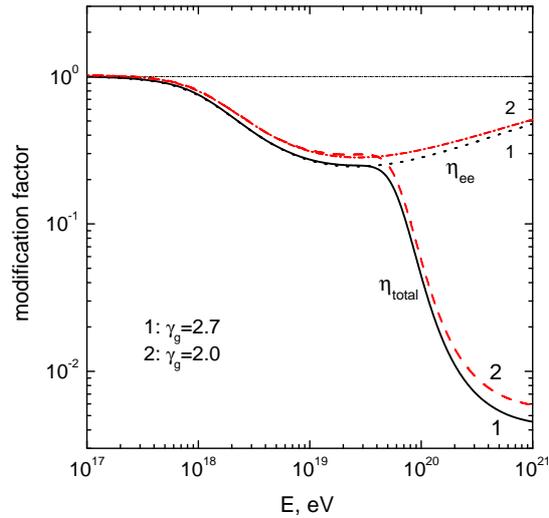} }
\caption{Modification factors for two generation indices $\gamma_g=2.7$
and 2.0. The dotted curve shows $\eta_{\rm ee}$ when only adiabatic and
pair-production energy losses are included. The solid and dashed
curves include also pion-production losses.
}
\label{fig:modfactor}
\end{center}
\end{figure}
Theoretical modification factors calculated for different source
generation indices $\gamma_g$ are presented in
Fig.~\ref{fig:modfactor}. If one includes in the calculation of
$J_p(E)$ only adiabatic energy losses, then, according to its
definition, $\eta(E)=1$ (dash-dot line in
Fig.~\ref{fig:modfactor}). When $e^+e^-$-production is
additionally included, one obtains spectrum $\eta(E)$ shown in
Fig.~\ref{fig:modfactor} by the curves labeled as $\eta_{ee}$.
With the pion photo-production process being also included, the
GZK feature (curves ``total'') appears. The observable part of the
dip extends from the beginning of the GZK cutoff at $E \approx
40$~EeV down to $E \approx 1$~EeV, where $\eta \approx 1$. It has
two flattenings: one at energy $E_a^{\rm tr} \sim 10$~EeV and the
other at $E_b \sim 1$~EeV. The former automatically produces the
ankle (see Fig.~\ref{fig:dips}) and the latter provides an
intersection of the flat extragalactic spectrum at $E \leq 1$~EeV
with the steeper Galactic one.

We discussed above the theoretical modification factor. The
{\em observed} modification factor, according to definition, is given by the
ratio of the observed flux $J_{\rm obs}(E)$ and unmodified spectrum
$J_{\rm unm}(E) \propto E^{-\gamma_g}$, defined up to normalization as:
$\eta_{\rm obs} \propto J_{\rm obs}(E)/E^{-\gamma_g}$.
Here $\gamma_g$ is the exponent of the generation spectrum $Q_{\rm
gen}(E_g) \propto E_g^{-\gamma_g}$ in terms of initial proton energies
$E_g$. Fig.~\ref{fig:dips} shows that both the pair production dip and
the beginning of the GZK cutoff up to $80$~EeV are well
confirmed by experimental data \cite{data} of Akeno-AGASA, HiRes,
Yakutsk and TA.
\begin{figure}
\begin{center}
 \begin{minipage}[h]{53mm}
 \includegraphics[width=53mm,height=51mm]{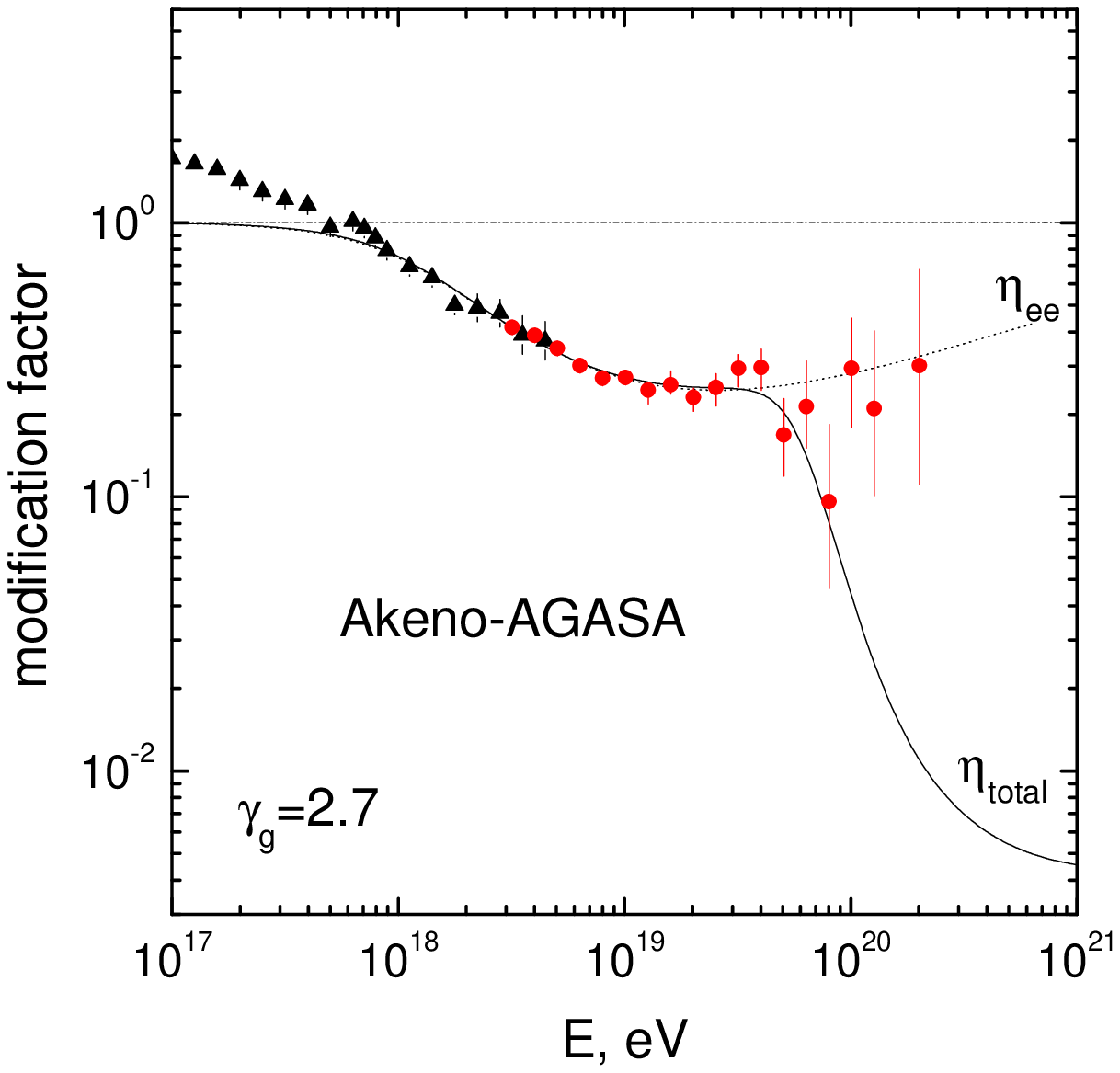}
 \end{minipage}
 \hspace{1mm}
 \begin{minipage}[h]{54 mm}
 \includegraphics[width=54mm,height=51mm]{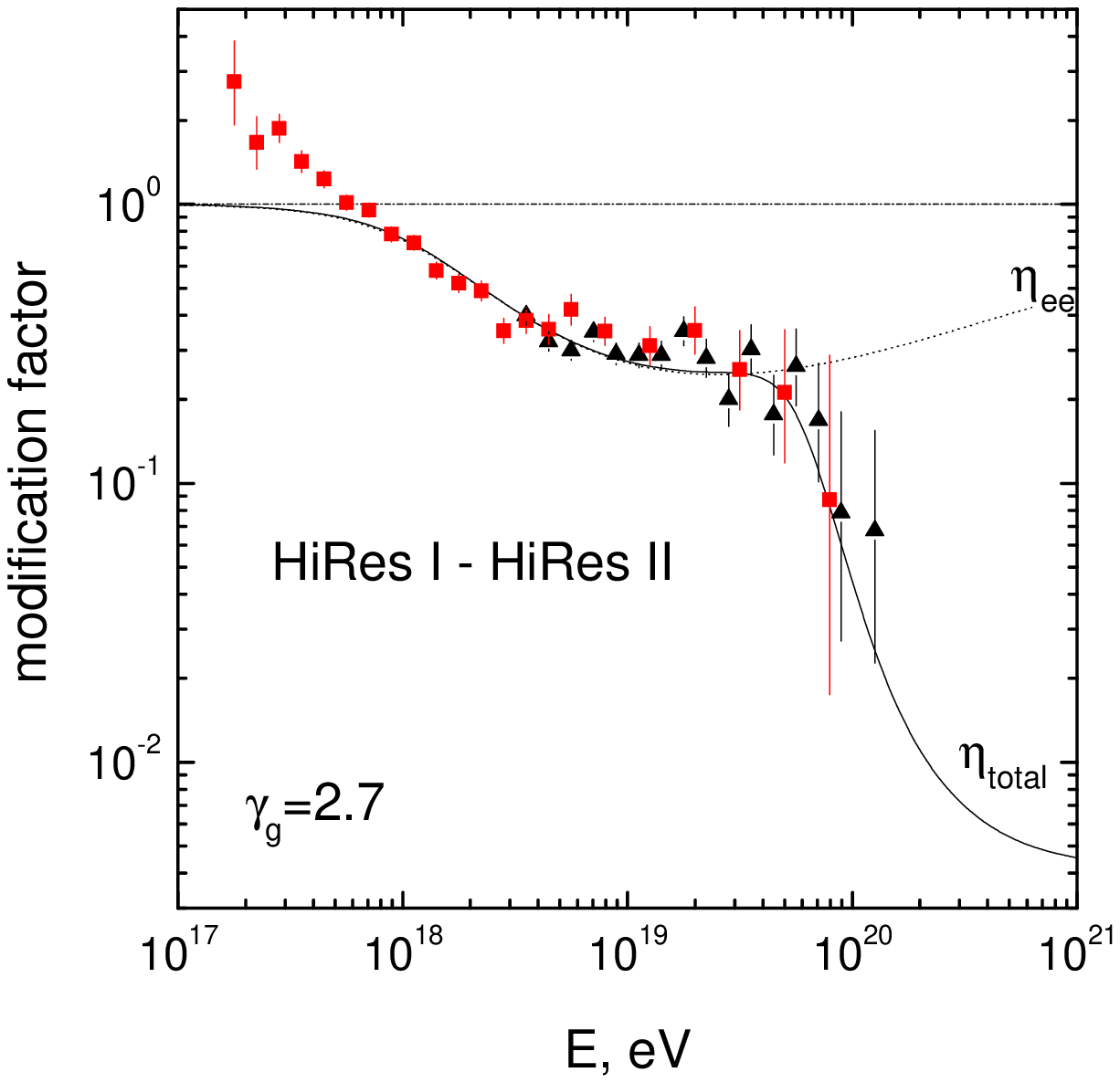}
 \end{minipage}
\newline \noindent
\medskip \hspace{-8mm}
 \begin{minipage}[ht]{54mm}
 \includegraphics[width=53mm,height=53mm]{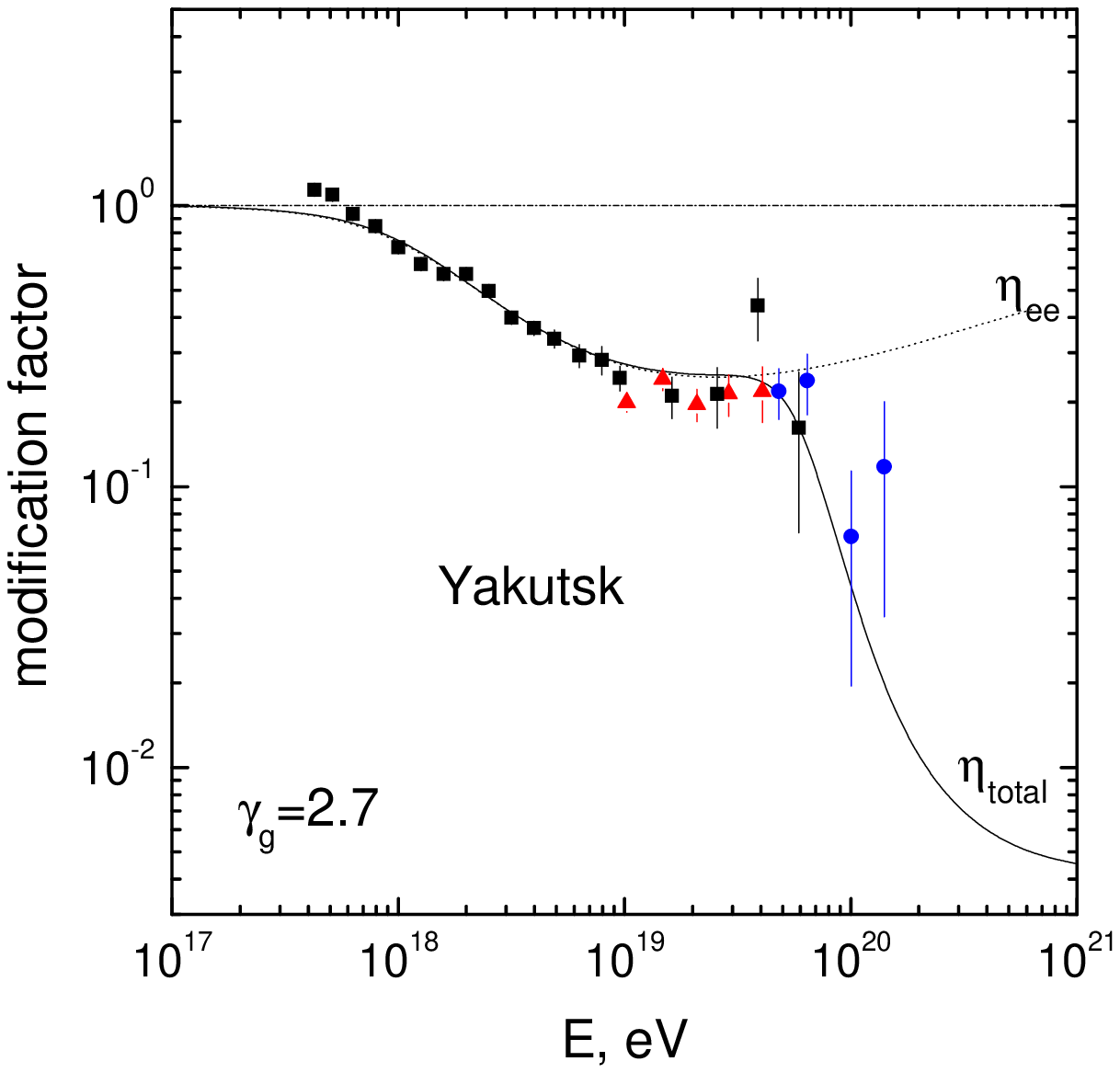}
 \end{minipage}
 \hspace{2mm}
 \begin{minipage}[h]{54 mm}\medskip
 \includegraphics[width=53mm,height=54mm]{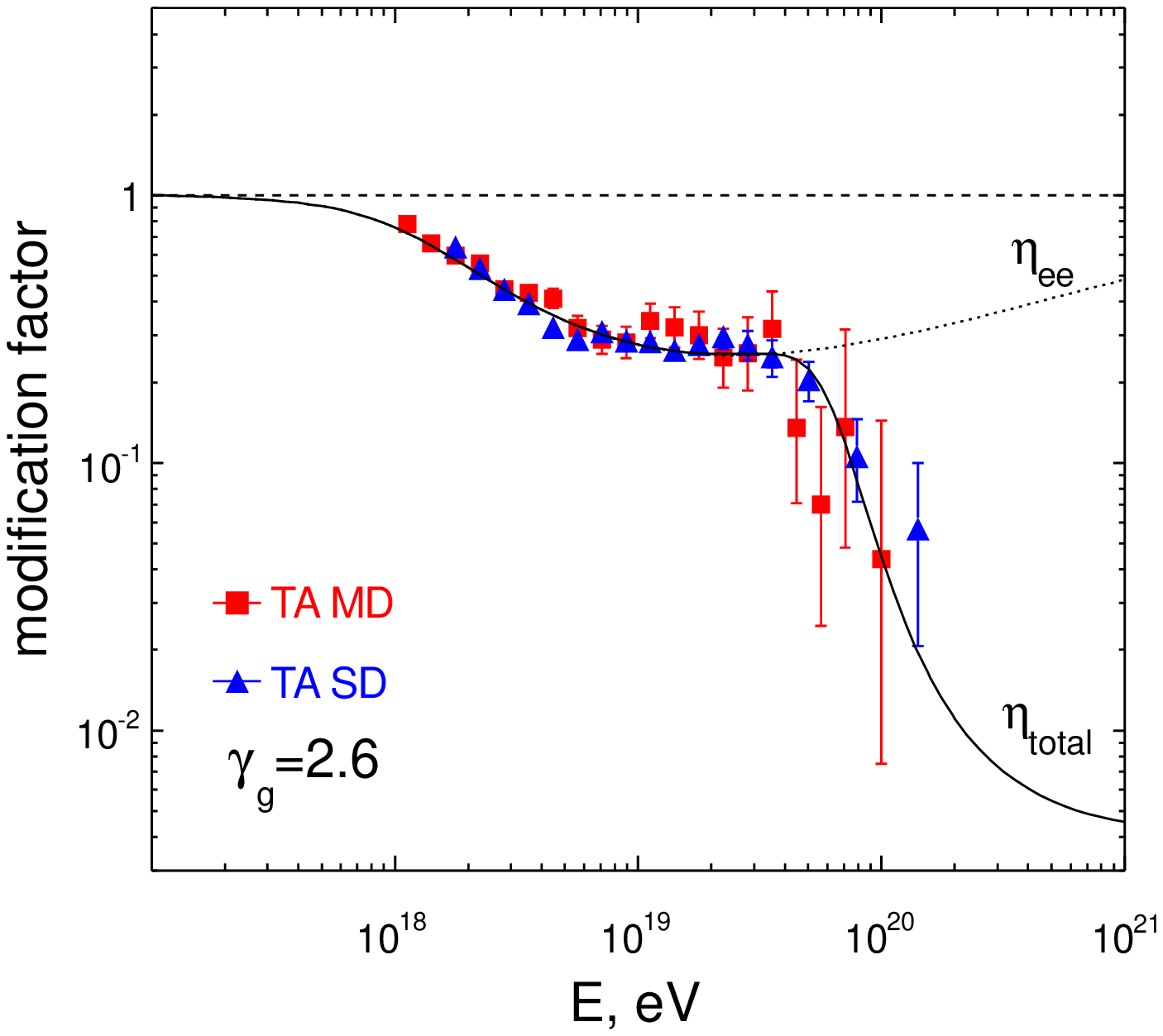}
\end{minipage}
 \vspace{-2 mm}%
\caption{ The predicted pair-production dip in comparison with
Akeno-AGASA, HiRes, Yakutsk and Telescope Array data \cite{data}.
All these experiments confirm the dip behavior with good accuracy,
including also the data of Fly's Eye \cite{data} (not presented
here).
} %
\label{fig:dips}
\end{center}
\end{figure} %
The comparison of the theoretical dip with observational data includes
only two free parameters: exponent of the power-law generation spectrum
$\gamma_g$ (the best fit corresponds to $\gamma_g=2.6 - 2.7$) and the
normalization constant to fit the $e^+e^-$-production dip to the
measured flux. The number of energy bins in the different experiments is
$20 - 22$. The fit is characterized by $\chi^2/{\rm d.o.f.} = 1.0 - 1.2$
for AGASA, HiRes and Yakutsk data.  This is a very good fit for {\em
signature} (see beginning of this section). For this fit we used the
modification factor without cosmological evolution of sources. As was
explained above, using a model approach with additional three
parameters describing the cosmological evolution one can further
improve the agreement.
In Fig.~\ref{fig:dips} one can see that at $E \lesssim 0.6$~EeV the
experimental modification factor, as measured by Akeno and HiRes,
exceeds the theoretical modification factor. Since by definition the
modification factor must be less than one, this excess signals the
appearance of a new component of cosmic rays at $E < E_{\rm tr} \approx
0.6$~EeV, which can be nothing else but the Galactic cosmic rays. This
interpretation is confirmed by transition of heavy component to the
protons in the upper-left panel of Fig.~\ref{fig:GZK-hires}, 
that with good accuracy occurrs at the
same energy. Thus, according to HiRes data the transition from
extragalactic to Galactic cosmic rays, occurs at energy $E_{\rm tr}
\sim 0.6$~EeV and is accomplished at $E \sim E_b \approx 1$~EeV (see
upper-left panel in Fig.~\ref{fig:GZK-hires} as example).

\subsection{Pair-production dip as energy calibrator}
\label{sec:energy-calibrator}
The energy position of pair-production dip is rigidly fixed by interaction
with CMB and thus it can serve as energy calibrator for the detectors.

As we already mentioned, it is difficult to expect that in terms of
the modification factor the dip described by one free physical
parameter can fit the observational data with minimum $\chi^2$.
One can shift the observed energy bins by the recalibration factor
$\lambda_{\rm cal}$, within the systematic error of observations,
to minimize $\chi^2$ \cite{BGGprd}.
We shall refer to this procedure as 'recalibration of energy scale'.

\begin{figure}\medskip
\begin{center}
 \begin{minipage}[h]{60mm}
 \includegraphics[width=60mm,height=47mm]{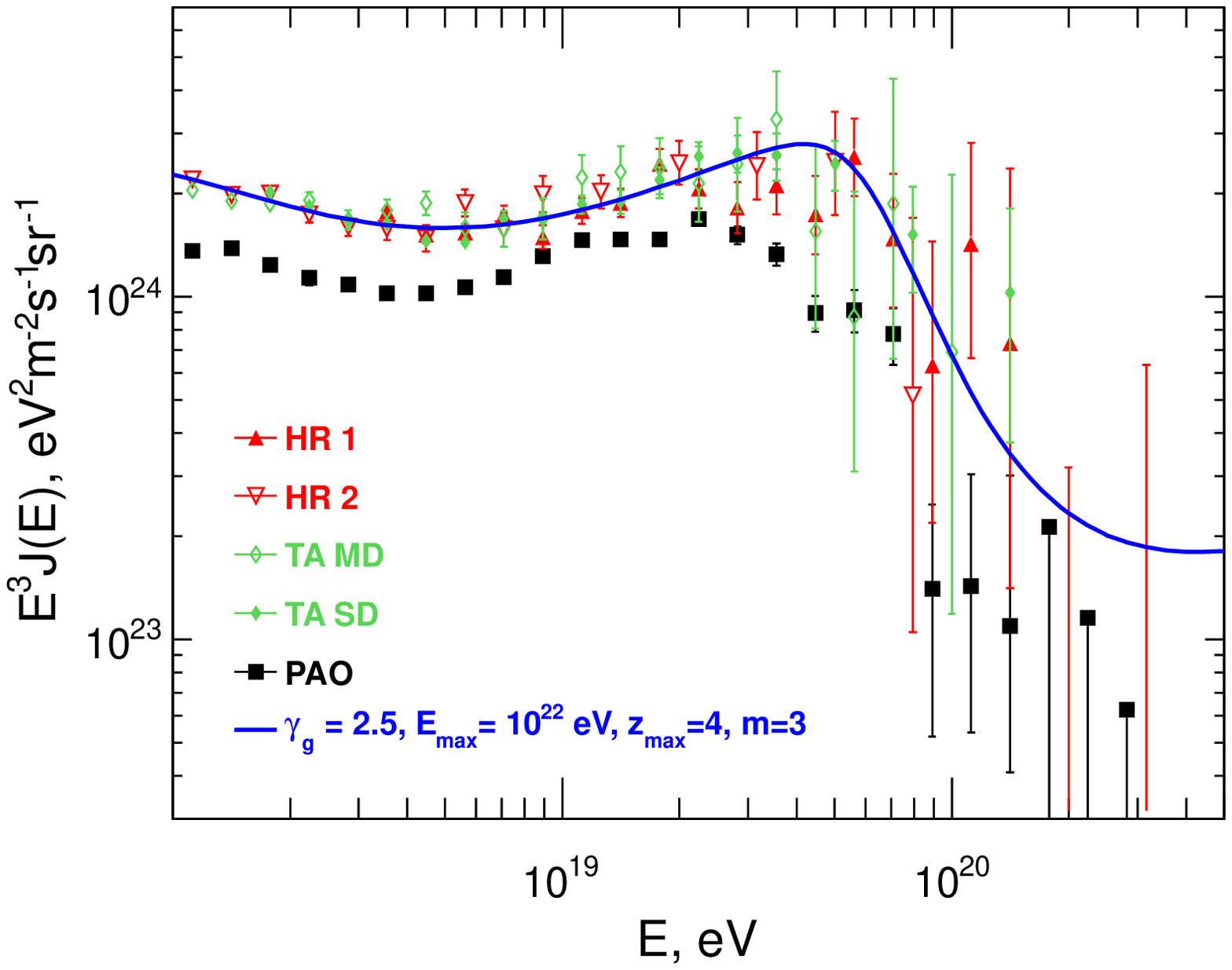}
 \end{minipage}
 \hspace{4mm}
 \begin{minipage}[h]{60mm}
 \includegraphics[width=60mm,height=47mm]{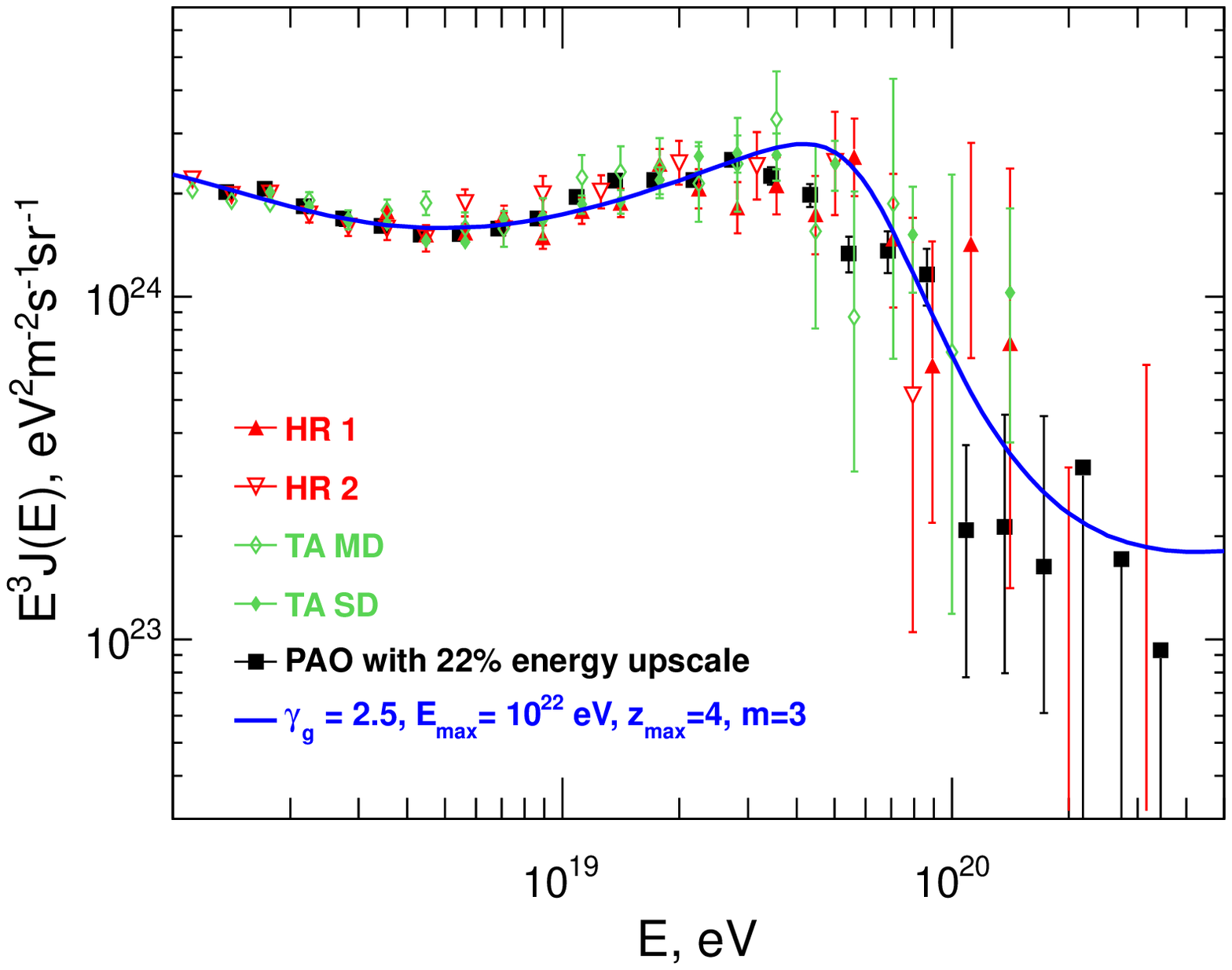}
 \end{minipage}
\caption{{\em Left panel:} Comparison of the PAO energy spectrum
(filled boxes) with the HiRes and TA data fitted by theoretical
pair-production dip (solid curve). {\em Right panel:} Spectra
after energy recalibration of the PAO data with $\lambda_{\rm
cal}=1.22$ (see the text). } \label{fig:HiTApao-rec}
\end{center}
\end{figure}

\begin{figure}\medskip
\begin{center}
 \begin{minipage}[h]{60mm}\hspace{-5mm}
 \includegraphics[width=60mm,height=47mm]{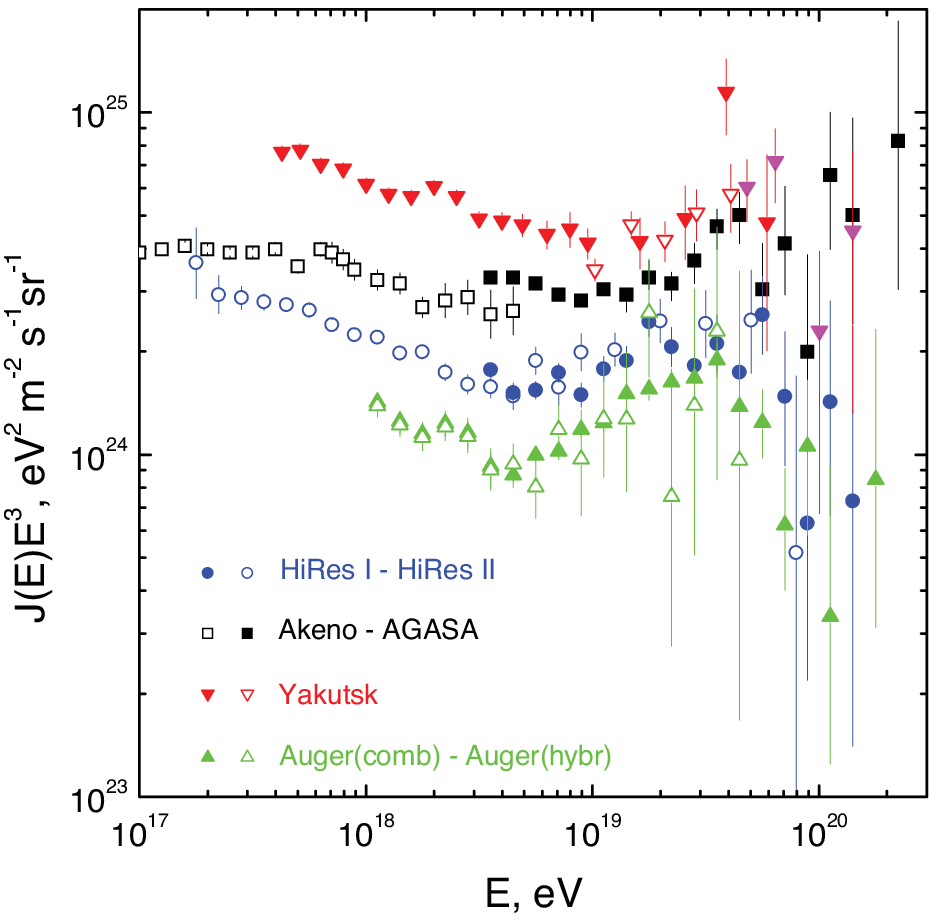}
 \end{minipage}
 \hspace{4mm}
 \begin{minipage}[h]{60mm}\vspace{3mm}\hspace{-5mm}
 \includegraphics[width=60mm,height=45mm]{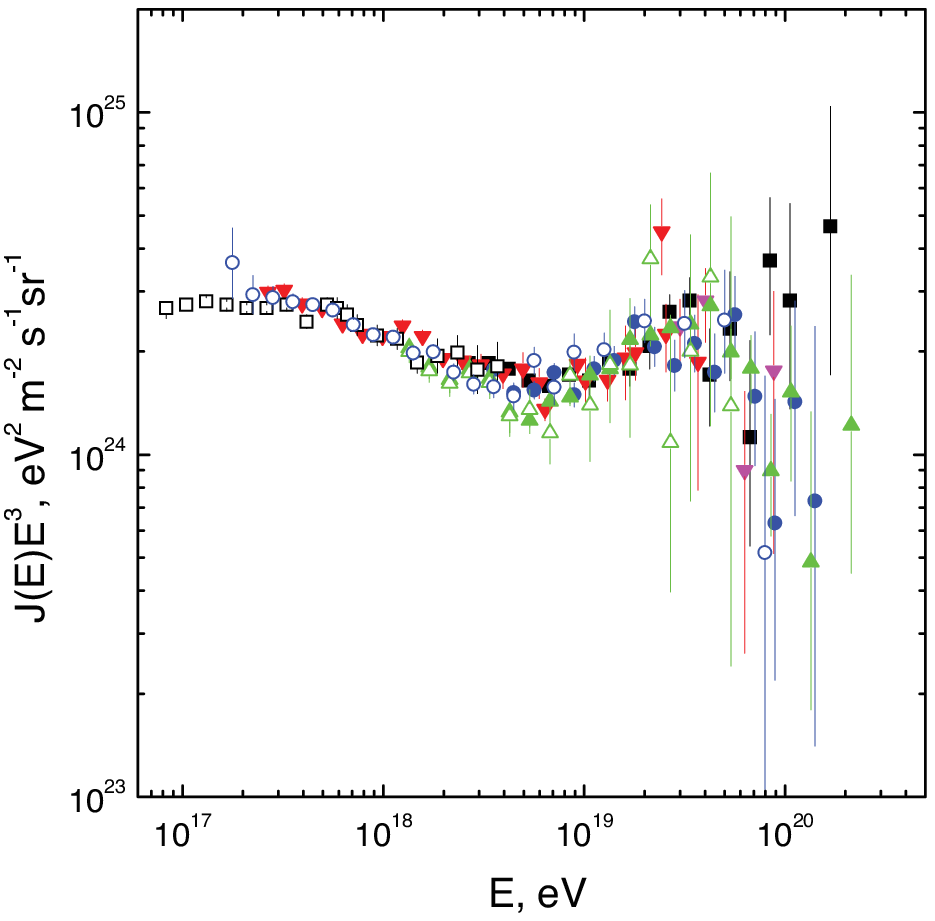}
 \end{minipage}
\caption{{\em Left panel:} Original fluxes from all detectors
(fluxes from HiRes and TA are approximately the same). {\em Right
panel:} Spectra after energy recalibration by pair-production dip:
$\lambda_{\rm cal}=1$ for HiRes/TA, $\lambda_{\rm cal}=1.22$ for
PAO, $\lambda_{\rm cal}=0.75$ for AGASA and $\lambda_{\rm
cal}=0.625$ for Yakutsk. } \label{fig:recalib-all}
\end{center}
\end{figure}

We discuss first the dip in PAO spectrum \cite{auger-spectrum}
presented by the filled boxes in the left panel of
Fig.~\ref{fig:HiTApao-rec}. As it was already mentioned, because of
very small statistical error bars it has too large $\chi^2$ in
comparison with pair-production dip terms of modification factor.
We shall use then the model-dependent method in terms of $E^3J(E)$
including the cosmological evolution $(1+z)^m$ up to $z_{\max}$
as shown in the left panel of Fig.~\ref{fig:HiTApao-rec}. Using two
more free parameters $m$ and $z_{\max}$ we can reach better agreement
with the modified shape of the dip shown the solid curve in
Fig.~\ref{fig:HiTApao-rec}. Now we can shift the PAO energy bins by
factor $\lambda_{\rm cal}$ reaching the minimum $\chi^2$. For this
$\lambda_{\rm cal}=1.22$ is needed. As a result we obtain picture shown in
the right panel of  Fig.~\ref{fig:HiTApao-rec}. We obtained not only
the excellent agreement with the shape of the theoretical
pair-production dip (solid curve) but also the good agreement with
absolute fluxes of HiRes and TA. Note that disagreement with GZK
cutoff remains for the three energy bins in energy interval $35 - 52$~EeV.  %
Recalibration with help of pair-production dip for all five detectors
(HiRes, Telescope Array, PAO, AGASA and Yakutsk) is shown in
Fig.~\ref{fig:recalib-all}. Recalibration factor $\lambda_{\rm cal} =1$
for HiRes/TA is based on the scale factor which correctly describes
the pair-production dip and  GZK cutoff in differential and integral
($E_{1/2}$) spectra.
\subsection{GZK cutoff in HiRes and Telescope Array data}
\label{sec:GZK}
The two largest Extensive Air Shower (EAS) detectors, HiRes \cite{GZK-hires}
and Pierre Auger Observatory \cite{GZK-PAO} have observed a sharp
steepening in the UHECR spectrum at $E \gtrsim (30 - 50)$~EeV. Both
collaborations claimed that the observed steepening is consistent with the
GZK cutoff. But as a matter of fact, there is a dramatic conflict between
these two results, which still leaves the problem open.
\begin{figure}
\begin{center}
 \begin{minipage}[h]{60mm}
 \includegraphics[width=60mm,height=45mm]{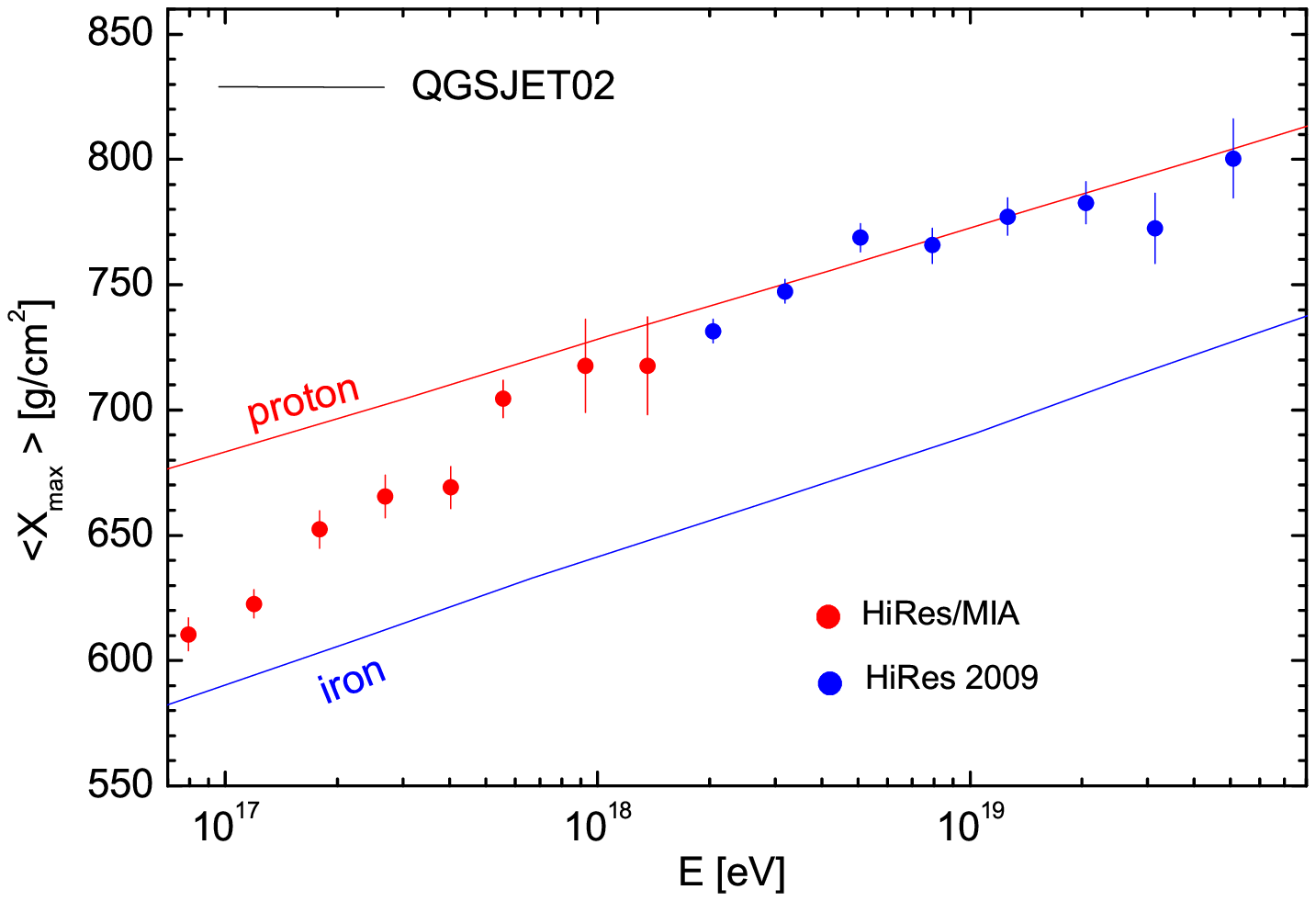}
 \end{minipage}
 \hspace{3mm}
 \begin{minipage}[h]{60mm}
 \includegraphics[width=60mm,height=45mm]{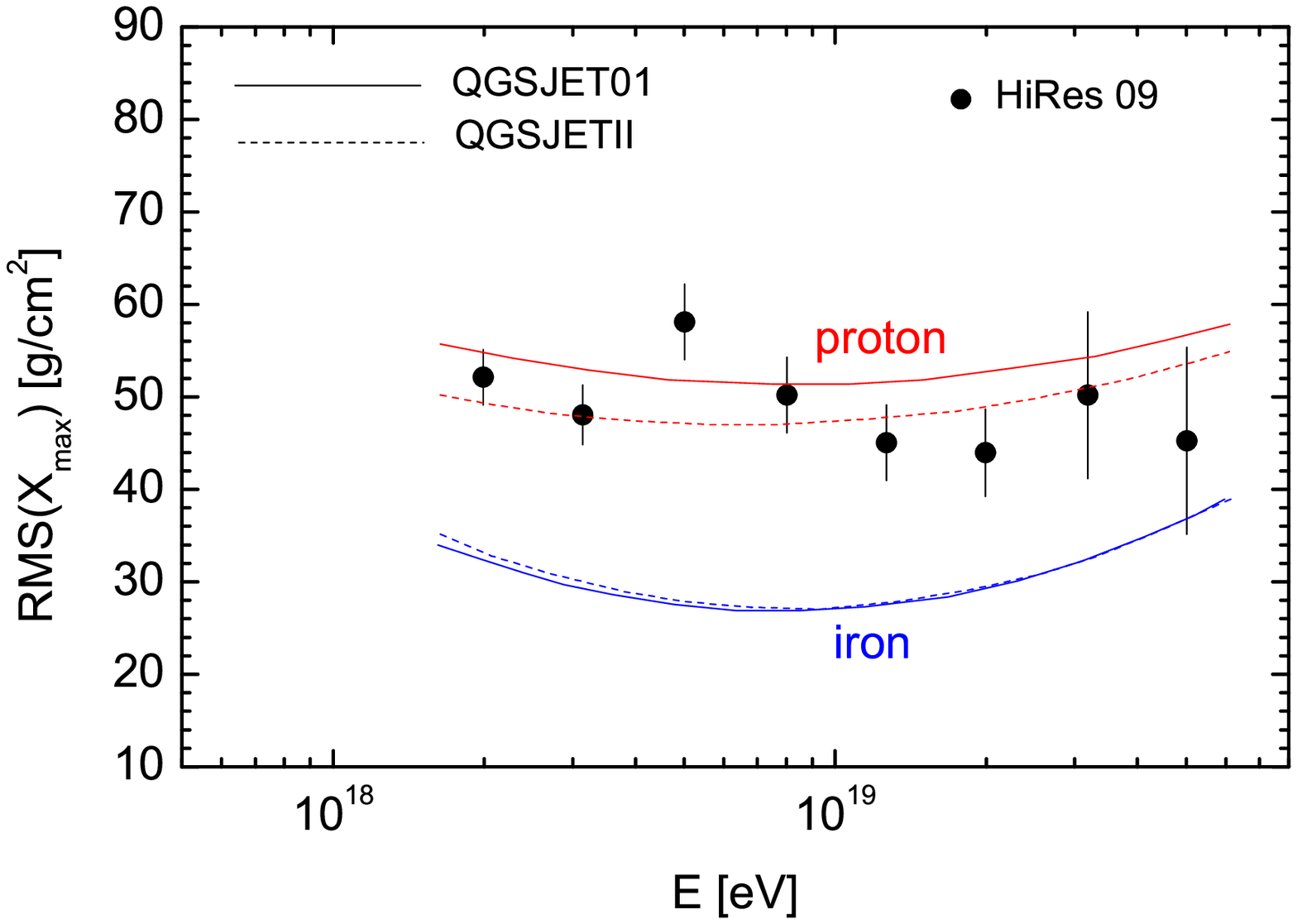}
 \end{minipage}
\newline \noindent
\medskip \hspace{-8mm}
 \begin{minipage}[ht]{66mm}\vspace{2mm}
 \includegraphics[width=66mm,height=45mm]{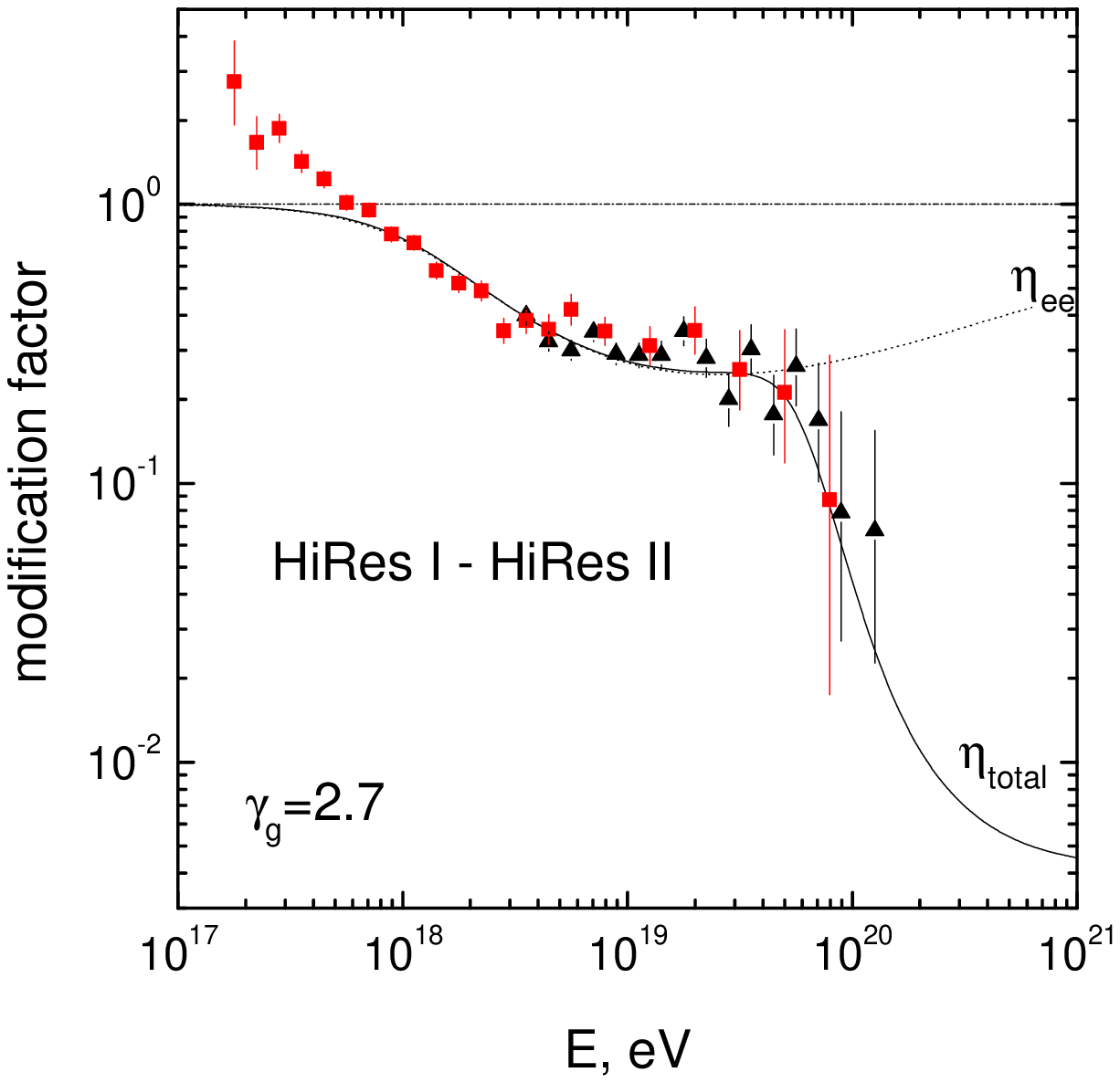}
 \end{minipage}
 \hspace{3mm}
 \begin{minipage}[h]{60mm}\medskip
 \includegraphics[width=59mm,height=45mm]{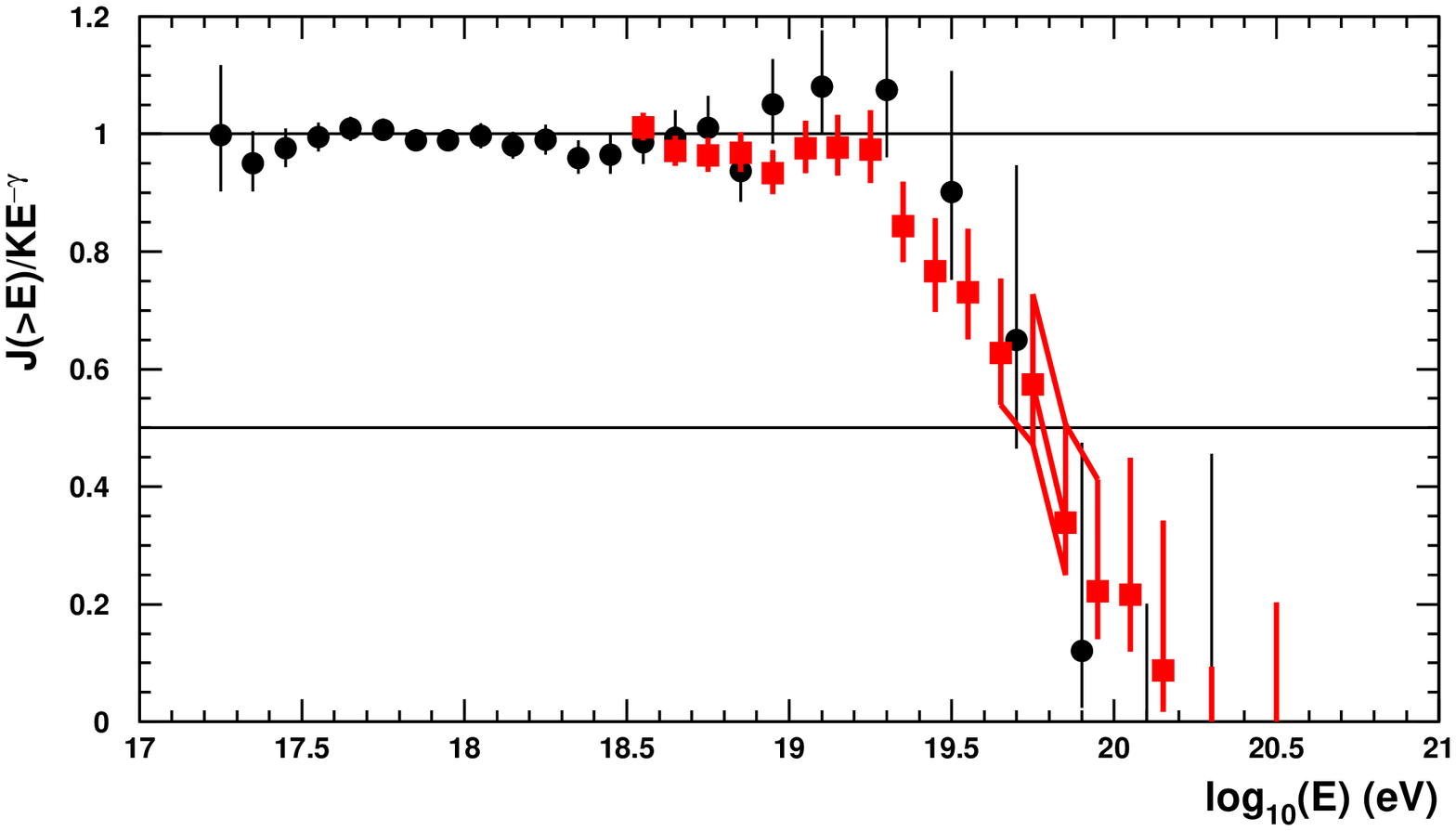}
\end{minipage}
 \vspace{-2 mm}%
\caption{ Mass composition and GZK cutoff as measured by the HiRes
detector. In two upper panels $\left\langle X_{\max}\right\rangle$
(left) and RMS (right) are presented as function of the energy.
Both agree with a pure proton composition, shown by curves labeled
'proton'. The left-lower panel shows differential energy spectrum
in terms of the modification factor. One can see a good agreement
with the predicted shape of the GZK cutoff. The right-lower panel
shows the quantity $E_{1/2}$ in the integral spectrum. This
energy, a characteristic of the GZK cutoff, is found as $E_{1/2}=
10^{19.73\pm 0.07}$~eV in good agreement with theoretical
prediction $E_{1/2}= 10^{19.72}$~eV (see the text).
} %
\label{fig:GZK-hires}
\end{center}
\end{figure} %
In this subsection we analyze data of the HiRes which provide a
strong evidence in favour of the GZK cutoff. These data are
supported also by the TA data \cite{GZK-TA}. The data of PAO will
be considered in the next subsection.

To interpret convincingly the spectrum steepening as the GZK cutoff
one must prove that ({\em i}) energy scale of the cutoff and its shape
correspond to theoretical predictions and ({\em ii}) the measured
mass composition is strongly dominated by protons. In HiRes the mass
composition is determined from $\left\langle X_{\rm max}\right\rangle(E)$,
average depth of atmosphere in $g/cm^2$, where a shower with energy $E$
reaches maximum, and RMS($X_{\max})$, which is the width of the distribution
over $X_{\max}$. These values measured by the HiRes are displayed in
Fig.~\ref{fig:GZK-hires}. From the left-upper panel of Fig.~\ref{fig:GZK-hires}
one can see that the chemical composition changes from very heavy elements,
probably Iron, at $E \sim 0.1$~EeV, (data of HiRes-MIA \cite{hires-mia}) to
protons at $E\sim 1$~EeV (data of HiRes \cite{ankle-hires}). RMS$(X_{\max})$,
a very sensitive tool for mass composition, also provides evidence for a
proton-dominated composition at $E \gtrsim 1$~EeV and up to the highest
energies (see upper-right panel of Fig.~\ref{fig:GZK-hires}). Differential
energy spectrum of the GZK feature in the form of modification factor
(left-lower panel) is in a reasonably good agreement with the theoretical
prediction, though better statistics at higher energies is still needed for
a final conclusion.

The {\em integral energy spectrum} of UHE protons, $J_p(>E)$, has
another specific characteristic of the GZK cutoff, the energy $E_{1/2}$
\cite{BG88}. It is based on the observation that the calculated
integral spectrum below $50$~EeV is well approximated by a power-law
function: $J_p(>E) \propto E^{-\tilde{\gamma}}$. At high energy this
spectrum is steepening due to the GZK effect. The energy where this steep
part of the spectrum equals to the half of its power-law extrapolation,
$J_p(>E)=K E^{-\tilde{\gamma}}$, defines the value of $E_{1/2}$. This
quantity is found to be practically model-independent; it equals to
$E_{1/2} = 10^{19.72}\mbox{ eV} \approx 52.5 $ EeV \cite{BG88}.
Fig.~\ref{fig:GZK-hires} demonstrates how the HiRes collaboration found
$E_{1/2}$ from observational data \cite{E_1/2hires}. The ratio of the
measured integral spectrum $J(>E)$ and the low-energy power-law
approximation $KE^{-\tilde{\gamma}}$ was plotted as a function of
energy. This ratio is practically constant in the energy interval $0.3 -
40$~EeV, indicating that the power-law approximation is a good fit,
indeed. At higher energy the ratio falls down and intersects the
horizontal line $0.5$ at the energy defined as $E_{1/2}$. It results in
$E_{1/2}= 10^{19.73\pm 0.07}$~eV, in an excellent agreement with the
predicted value.

Thus, one may conclude that the HiRes data presented in
Fig.~\ref{fig:GZK-hires} indicate the proton-dominated chemical
composition and the presence of the GZK cutoff in both differential and
integral spectra. The conclusion about proton composition is further
supported by the recent TA data \cite{TAicrc11}.
\begin{figure}\medskip
\begin{center}
 \begin{minipage}[h]{60mm}
 \includegraphics[width=60mm,height=40mm]{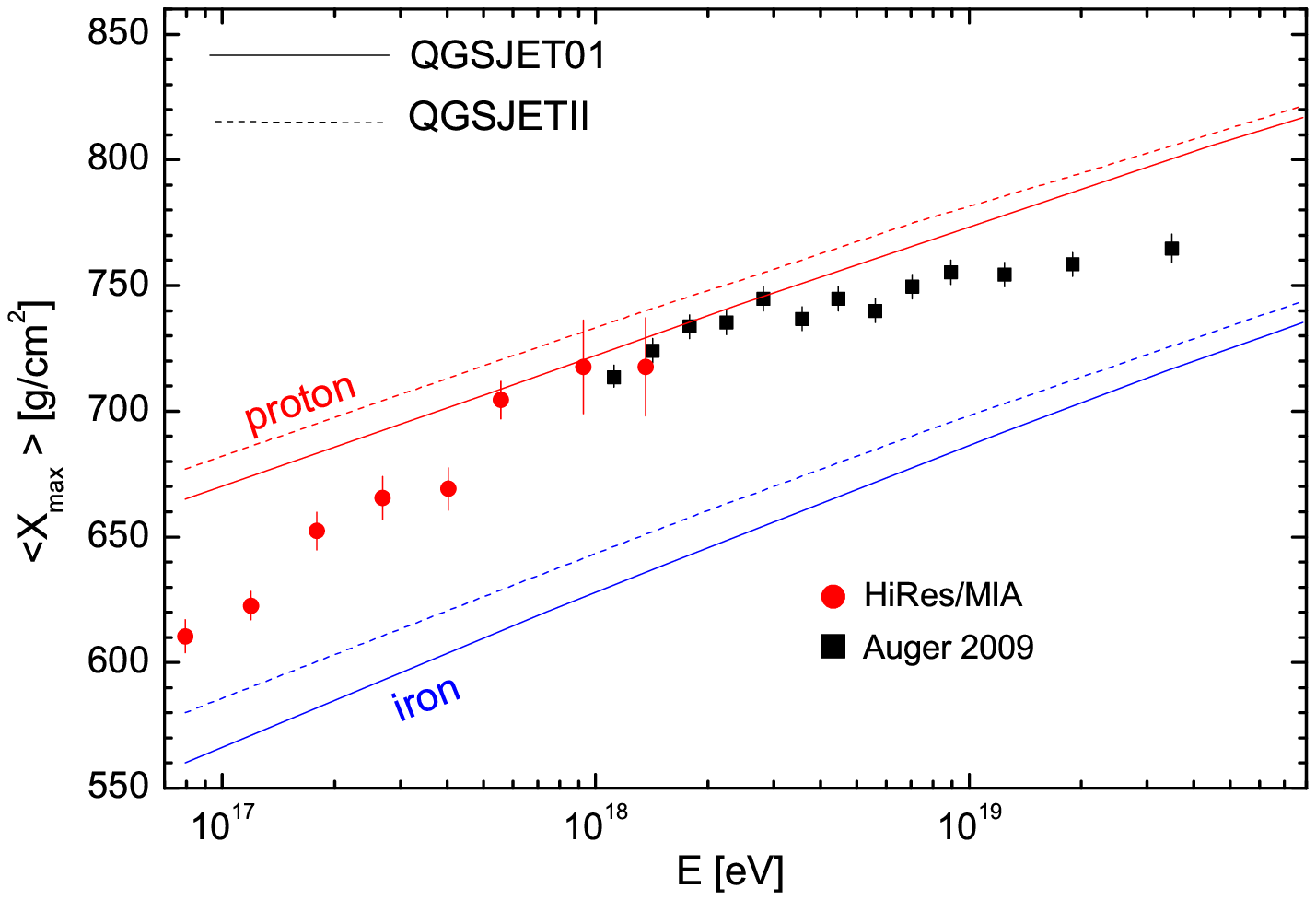}
 \end{minipage}
 \hspace{4mm}
 \begin{minipage}[h]{62mm}\vspace{1mm}\hspace{2mm}
 \includegraphics[width=61mm,height=43mm]{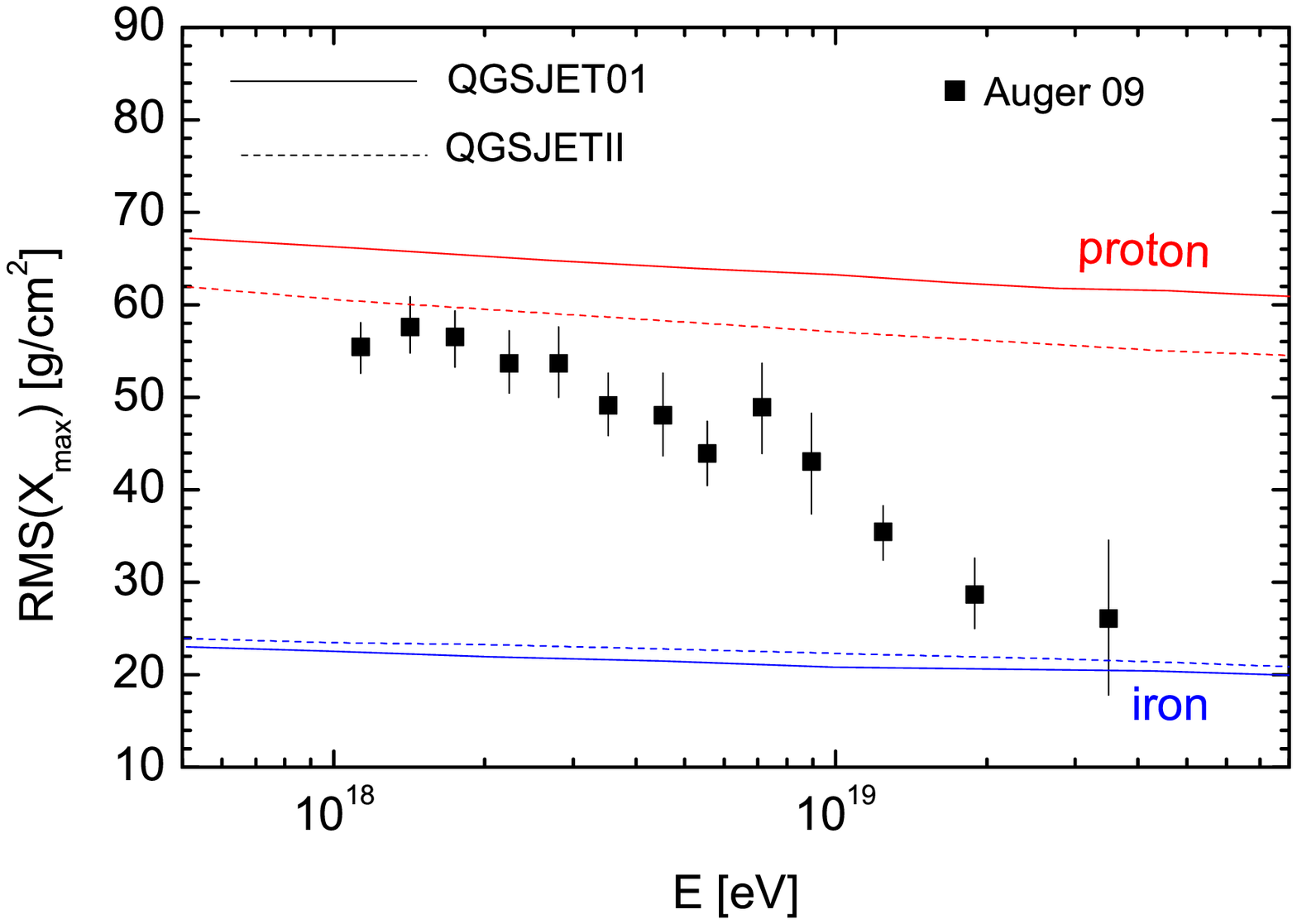}
 \end{minipage}
 \vspace{-2mm}
\caption{{\em Left panel:} Auger data \cite{auger-mass} for
$\left\langle X_{\max}\right\rangle$ as function of the energy
(left panel) and for RMS($X_{\max}$), the width of the
distribution over $X_{\max}$, (right panel). The calculated values
for protons and Iron are given according to QGSJET01 \cite{QGSJET}
and QGSJET II \cite{QGSJETII} models. One can see from the right
panel that RMS distribution becomes more narrow with increasing
energy which implies a progressively heavier mass composition. }
\label{fig:massAu}
\end{center}
\end{figure}
\subsection{PAO data: energy spectrum and mass composition}
\label{sec:auger}
In subsection~\ref{sec:energy-calibrator} we demonstrated that the
dip shape, as observed by PAO, can agree after recalibration with
energy spectra of HiRes/TA and other detectors (see
Figs.~\ref{fig:HiTApao-rec} and~\ref{fig:recalib-all} ). The coincidence
of the PAO and HiRes/TA  spectra is related to low energy part of the
energy spectrum in the right panel of Fig.~\ref{fig:HiTApao-rec}. At 
higher energies statistical uncertainties are too large to distinguish 
between the spectra.

While the HiRes and TA spectra are compatible with the GZK cutoff, the
Auger spectrum is not. The steepening in the upscaled PAO
spectrum  starts at energy $E \lesssim 40$ EeV, lower than $E_{\rm GZK}
\simeq 50$ EeV, and in three successive energy bins in the interval
$35 - 52$~EeV the PAO flux is significantly lower than one predicted
for the GZK shape as shown in the right panel of Fig.~\ref{fig:HiTApao-rec}.
We could not reconcile the PAO cutoff shape with the GZK behavior by
including in the calculations different generation indices $\gamma_g$,
evolution regimes, low acceleration maximum energy $E_{\max}$, local
overdensity of sources etc.

This disagreement is quite natural for the PAO mass composition
data which, in contrast to HiRes and TA, show strong dominance of
nuclei (see Fig.~\ref{fig:massAu}). A steepening in the end of the
of the nuclei spectrum, as calculations show, is quite different
from that of protons (GZK cutoff). The most reliable data on mass
composition is given by elongation curve $X_{\max}(E)$ and
especially by RMS($X_{\max}(E))$. In the Auger data the latter
steadily decreases with energy and approaches the Iron value at $E
\approx 35$~EeV. Low RMS, i.e.\ small fluctuations, is a typical
and reliable feature of the heavy nuclei composition. These data
are further strengthened by other PAO measurements provided by
surface detectors. They allow to extract two other
mass-composition dependent quantities: the atmospheric depth
$\left\langle X_{\max}^{\mu}\right\rangle$, where muon-production
rate reaches maximum, and maximum zenith angle $\theta_{\max}$
determined by the signal rise-time in surface Cerenkov detectors.
Measurements of both quantities confirm the heavy mass composition
and its dependence on energy obtained with the help of
$\left\langle X_{\max}\right\rangle(E)$ and RMS($X_{\max}$),
\cite{Garcia-GamezICRC} and \cite{Gazon2012}, respectively. The
soon expected data on muon flux from the Auger Muon and Infill
Ground Array (AMIGA) \cite{AMIGA} will further clarify the mass
composition.

Our further analysis of the Auger spectrum and mass composition is
based on the following two observations:\\*[1mm]
{\em (i)} According to the HiRes (Fig.~\ref{fig:GZK-hires}) and PAO
(Fig.~\ref{fig:massAu}) data, the observed primaries at energy
$(1 - 3)$~EeV are predominantly protons or nuclei not heavier than
Helium.\\*[1mm]
These particles cannot be galactic, otherwise, as MC simulations
\cite{anisotropy} show, galactic anisotropy would be too large.
Then the ankle in the PAO data is not a transition from galactic to
extragalactic cosmic rays, but transition from extragalactic
protons/Helium to extragalactic nuclei. Transition from galactic to
extragalactic CRs occurs thus at lower energies, most probably at the
second knee.\\*[1mm]
{\em (ii)}  The particles at higher energies are extragalactic nuclei
with the charge number $Z$ increasing with energy. \\*[1mm]
This observation is naturally explained by rigidity dependent
acceleration in the sources $E_i^{\max}=Z_i \times E_p^{\max}$,
since at each energy $E=ZE_p^{\max}$ the contribution of nuclei
with smaller $Z' < Z$ vanishes.

\begin{figure}
\begin{center}
\resizebox{0.4\textwidth}{!}{%
 \includegraphics{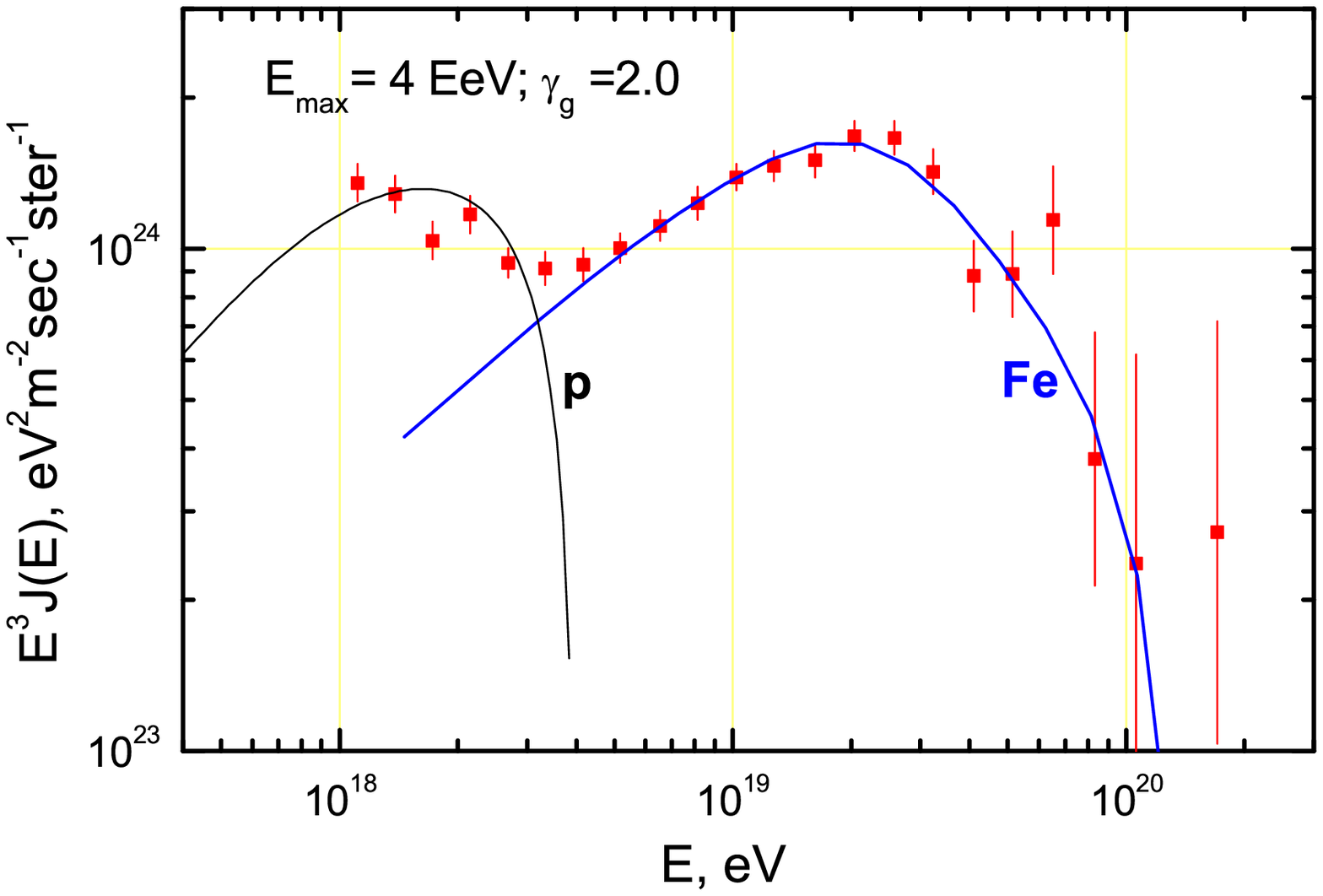} }
\resizebox{0.4\textwidth}{!}{%
 \includegraphics{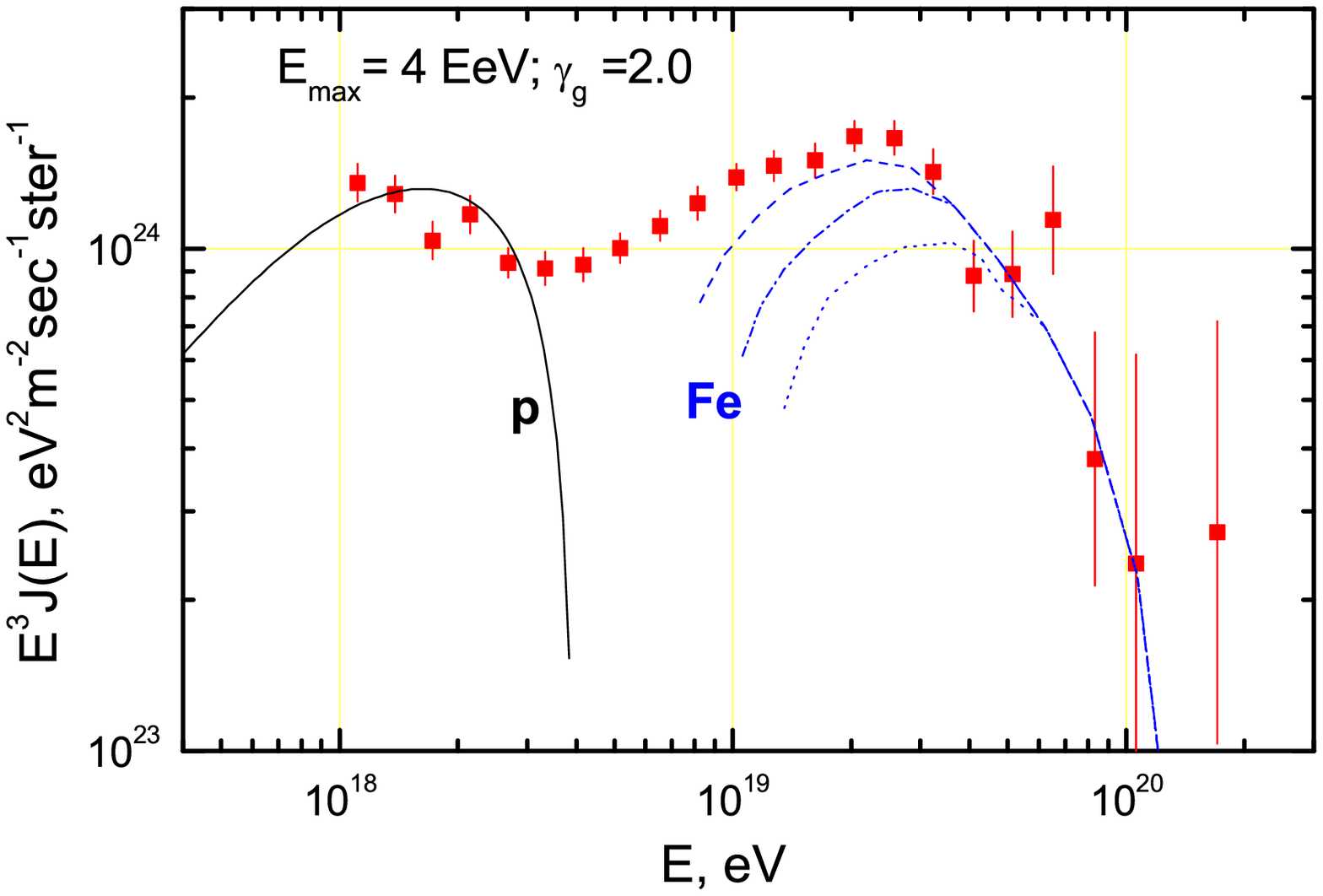} }
\caption{{\em Left panel:} Energy spectrum in the two-component model with
protons and Iron nuclei for homogeneous distribution of sources and
with $\gamma_g=2.0$ and $E_{\max}=4$~EeV. {\em Right panel:} As in
left panel but with a diffusion cutoff. The gap is expected to be filled
by intermediate mass nuclei.
}
\label{fig:disapp-spectrum}
\end{center}
\end{figure}

It was demonstrated in \cite{disapp} that to avoid a proton dominance at
the highest energies, one must assume that the maximum energy of the
accelerated protons is limited, $E_p^{\max} \lesssim (4 - 10)$~EeV.
This conclusion is valid for a large range of generation indices,
$\gamma_g \sim 2.0 - 2.8$, and for a wide range of cosmological
evolution parameters. The calculated proton and nuclei energy spectra
for $\gamma_g=2.0$, $E_{\max}=Z\times 4$~EeV and without cosmological
evolution are shown in Fig.~\ref{fig:disapp-spectrum}. In the left panel
the two component model (protons and Iron) is presented. In the right
panel intermediate primary nuclei
are included in the framework of the diffusive propagation through
intergalactic magnetic fields (see \cite{disapp} for details).
However, it is a problem to
explain simultaneously both the spectrum and the mass composition
of the PAO.
This model is called {\em'disappointing'} because of lack of many
signatures predicted in proton-dominated models, such as cosmogenic
neutrino production and correlation of CR arrival directions with
distant sources.

The similar model is considered in \cite{Allard2011} (see Fig.~4 there).
Like in \cite{disapp} the proton component with $E_p^{\max}=4$~EeV is
introduced and rigidity-dependent acceleration is considered. The more
detailed calculations including the secondary nuclei are performed for
Iron-enriched source spectrum. The calculated spectrum agrees well
with that of PAO at $E\gsim 3$~EeV.

More detailed calculations are performed in \cite{Taylor2011}.
For Iron or Silicon as the accelerated nuclei, and very flat
generation spectrum with $\gamma_g < 2.0$, the energy spectrum and mass
composition of the produced nuclei are calculated for rectilinear and
diffusive propagation. In some cases the agreement with Auger data is
reached for energy spectrum and RMS(E) (agreement with elongation curve
is worse). The rigidity-dependent acceleration is not assumed and problem with
proton/Helium component at (1 - 3)~EeV is not discussed. The presence
of nearby sources is emphasized as a general feature.
\section{Three transition models}
\label{transition}
In this section we discuss three models of transition from Galactic to
extragalactic CRs: {\em ankle}, {\em dip} and {\em mixed composition}
models. One feature is common for all of them: the transition is
described as an intersection of a steep Galactic spectrum with a
flat extragalactic one. The agreement with the SM for GCR
is one more criterion which these models have to respect.
According to the Standard Model, the benchmark of the end of GCR is
given by the Iron knee at energy $E_{\rm Fe} \approx 80$~PeV and
at $E > E_{\rm Fe}$ it has an exponential cutoff.

Motivated by the interpretation of the ankle as the transition to
extragalactic CR at $E_a^{\rm tr} \sim (3 - 10)$~EeV, one has to assume
\cite{hillas2005,hillas2006,gaisser} an additional component of GCR
accelerated to energies much beyond the Iron knee.

Observational data which have the power either to confirm or to reject
each transition model include energy spectrum, elongation curve
$\left\langle X_{\max}\right\rangle(E)$, RMS($X_{\max}$) and anisotropy.
Below these models are  discussed in the historical succession of
appearance: ankle, dip and mixed composition.
\subsection{Ankle model}
\label{AnkleModel}
This is the traditional model based on the interpretation of the ankle
as the spectrum feature where transition occurs (see e.g.\
\cite{stanev2005} - \cite{waxman}). In fact, this is a very natural
model since transition occurs because the extragalactic component is
very hard. This component is assumed to have a pure proton composition
with a flat generation spectrum $Q_{\rm extr.p} \propto E^{-2}$ valid
for non-relativistic shock acceleration. Energy losses modify the
spectrum insignificantly at $E \lesssim 40$~EeV. The beginning of the
ankle at $E_a^{\rm obs} \sim 4$~EeV corresponds to the energy
where fluxes of Galactic and extragalactic CRs get equal. Thus, the
Galactic CRs should be presented by an additional component accelerated
up to energy by factor $30 - 40$ times higher than the maximum
energy in the Standard Model. In the majority of ankle models, e.g.\
\cite{hillas2005,hillas2006,WW2005,waxman}, the large fraction of the
observed cosmic rays has a Galactic origin at $E \gtrsim 10$~EeV.
To facilitate the acceleration problem one should assume a heavy-nuclei
composition of the new component.

\begin{figure}
\begin{center}
 \begin{minipage}[h]{62mm}
 \includegraphics[width=60mm,height=43mm]{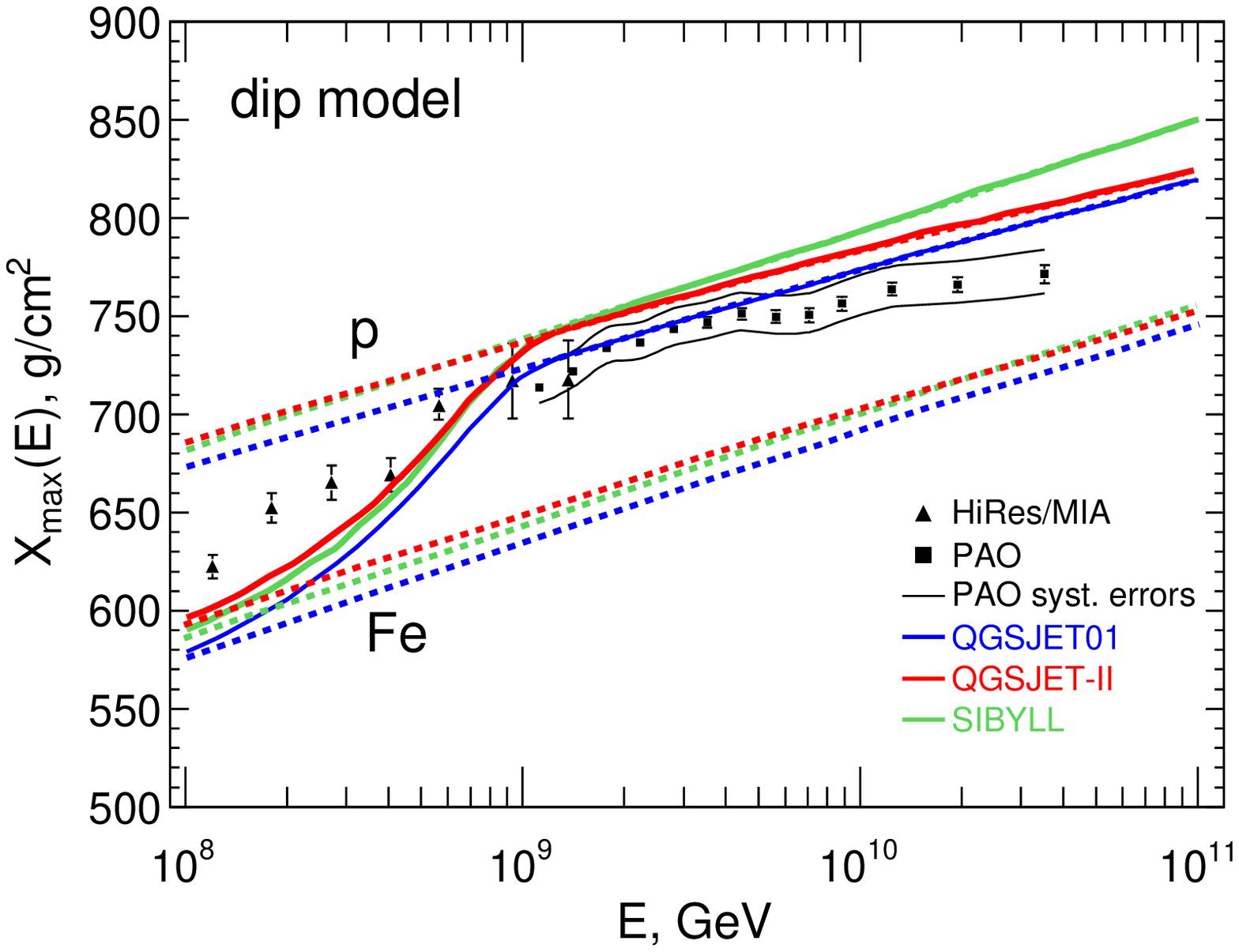}
 \end{minipage}
 \hspace{2mm}
 \begin{minipage}[h]{60mm}
 \includegraphics[width=60mm,height=43mm]{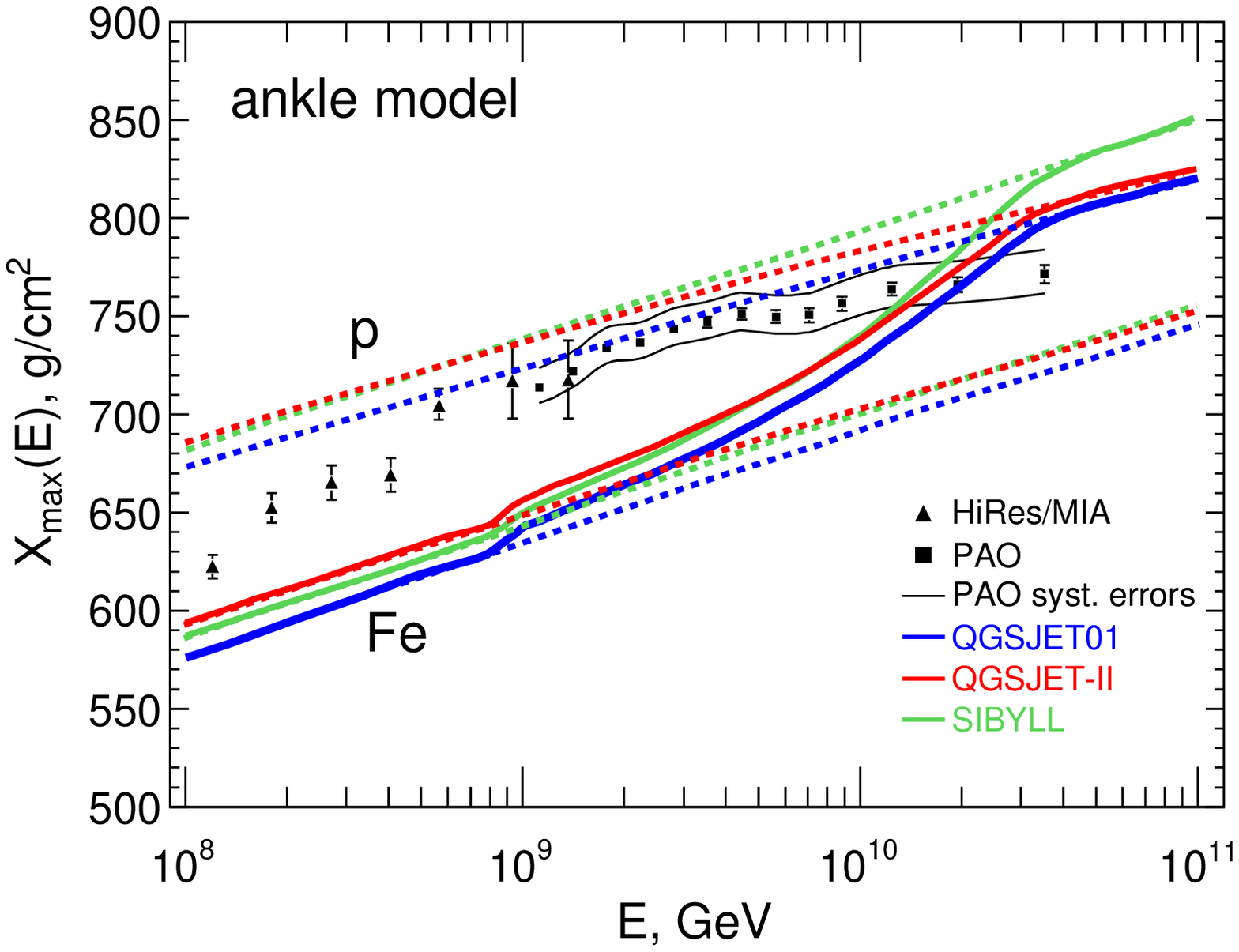}
 \end{minipage}
\newline \noindent
\bigskip \hspace{-5mm}
 \begin{minipage}[ht]{62mm} %\vspace{-1mm}
 \includegraphics[width=60mm,height=43mm]{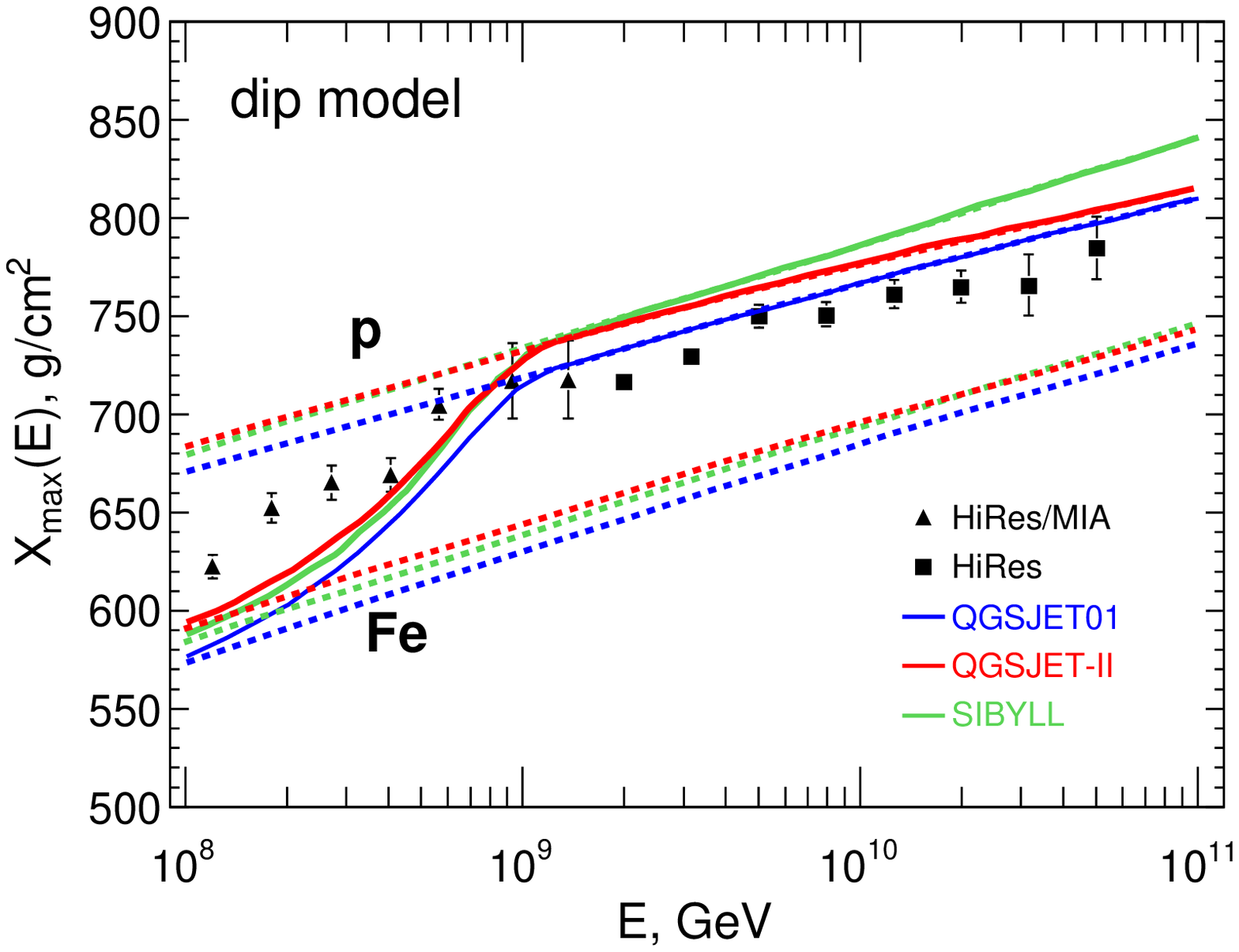}
 \end{minipage}
 \hspace{2mm}
 \begin{minipage}[h]{60mm} %\vspace{-1mm}
 \includegraphics[width=60mm,height=43mm]{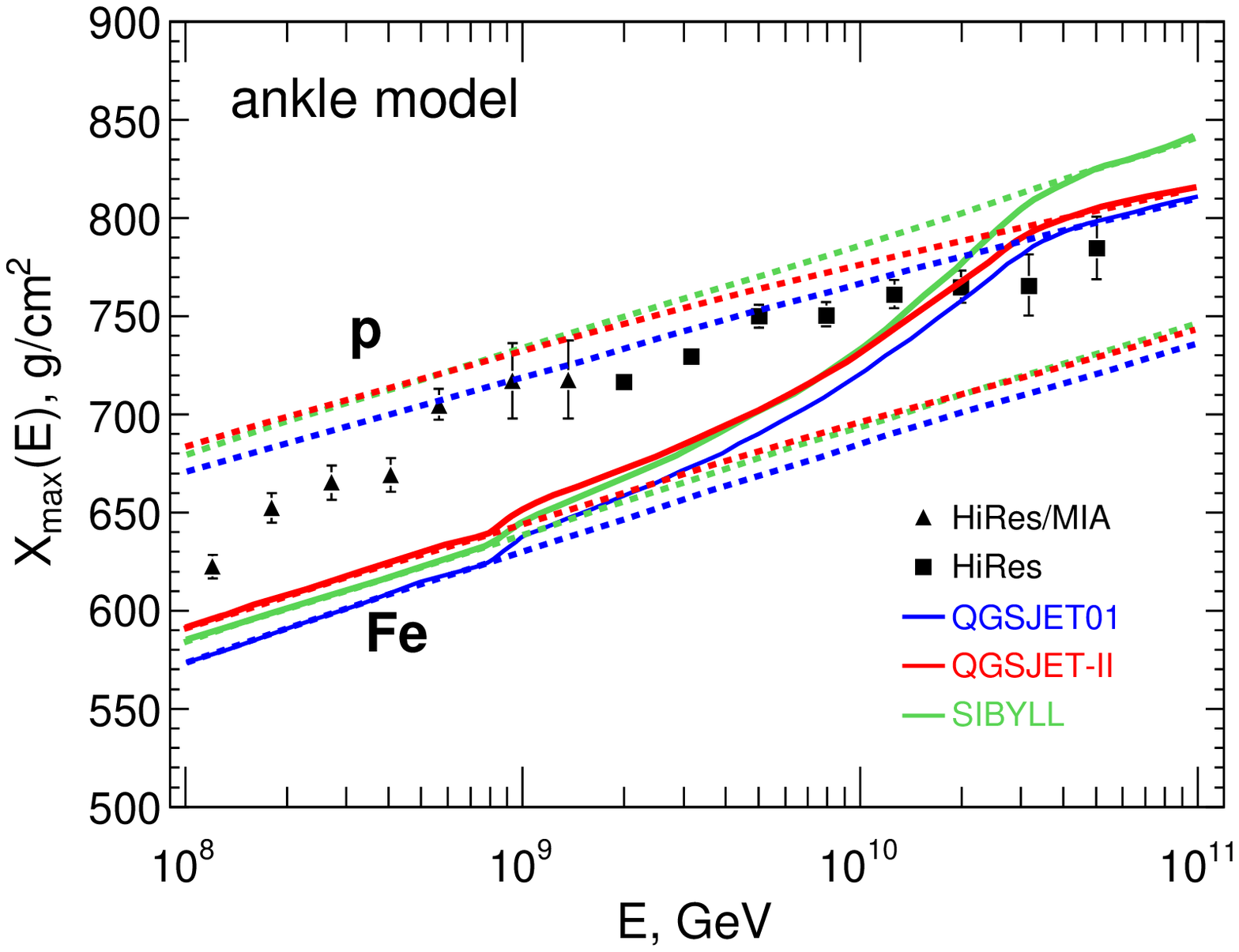}
\end{minipage}
 \vspace{-3 mm}%
\caption{ The calculated elongation curves $\left\langle
X_{\max}\right\rangle(E)$ for the dip model (two left panels) and
ankle model (right panels). The calculated  curves $\left\langle
X_{\max}\right\rangle(E)$ are shown by the thick solid lines for
QGSJET01 \cite{QGSJET}  model of interaction, by the thick dashed
lines for QGSJET-II \cite{QGSJETII}, and by dotted lines for
SIBYLL \cite{SIBYLL}. The data points are measurements of
HiRes-Mia (filled triangles), HiRes and PAO (both filled boxes).
The PAO data in upper panels with systematic errors, shown by the
thin curves, are taken from \cite{PAO-11cc}. The lines 'p' and
'Fe' present the elongation curves for proton and Iron which are
used for calculations of the model elongation curves in each
panel. For PAO data they are theoretical curves, for HiRes they
include cuts, detector's properties etc. and are taken from curves
'iron' and 'proton' in the upper-left panel of
Fig.~\ref{fig:GZK-hires}. As the main result of these plots one
may notice the great discrepancy of the ankle model with the data.
} \label{fig:Xmax}
\end{center}
\end{figure}

Another problem of the ankle model is the contradiction with the
measured average depth of EAS maximum, $\left\langle X_{\max}\right\rangle(E)$,
in the energy range $(1 - 4)$~EeV. While all data, including both HiRes and
PAO, show proton or light nuclei composition here, the ankle model
needs a heavy Galactic component, predicting too small
$\left\langle X_{\max}\right\rangle(E)$ in contradiction with observations
(see the right panels of Fig.~\ref{fig:Xmax} and right panel of Fig.~4 in
\cite{ABBO}). This contradiction is found also in \cite{anisotropy}
and \cite{mixed}.

Another contradiction in the ankle model is given by observed
proton/Helium composition at (1 - 3)~EeV. Since this energy
interval is below the ankle, i.e.\ has galactic origin, anisotropy
there should be too high \cite{anisotropy}. One may conclude that
ankle model is excluded or very strongly disfavoured.
%%%% Xmax  figure new %%%%%%%%%%%%%%%%%%%%%%%
\subsection{Dip model}
\label{sec:DipModel} The dip model is based on the assumption that
UHECRs at $E \gtrsim 1$ EeV are mostly extragalactic protons. This
assumption is confirmed by the HiRes and TA data (mass composition
and observation of pair-production dip and GZK cutoff), but
contradicts to PAO mass composition data. The transition from
galactic to extragalactic component begins at the second dip and
finishes at $E \sim 1$~EeV in good agreement with SM of Galactic
CRs.

The basic features of the {\em dip model} are as follows
\cite{BGGPL,BGGprd,Aletal}:\\
{\em (i)}~ The sources are AGN \cite{BGG-AGN} with a neutron
mechanism of particle escape \cite{berez77,dermerAGN} which provides a
pure proton generation spectrum.\\
{\em (ii)}~ The source generation spectrum has a usual shock-acceleration form
$q(E) \propto (E^{-2} - E^{-2.3})$, but generation rate per unit
comoving volume at high energy can be  $Q(E) \propto E^{-\gamma_g}$ with
$\gamma_g=2.6 - 2.7$ due to distribution of the sources over the
maximum energy of acceleration $n_s(E_{\max})$ \cite{KS}.\\
{\em (iii)}~ The generation index $\gamma_g=2.6 - 2.7$ is the main
fitting parameter (see subsection \ref{sec:modfactor-dip}).
The cosmological evolution
of the sources can be easily included, affecting mostly the low energy
part of the spectrum. Inclusion of additional parameters allows to
improve the fit. In particular, in \cite{BGGprd} the spectrum
was calculated with an account for cosmological evolution of AGN as it
follows from X-ray observations.\\
The confirmation of the dip model follows from: {\em (i)} agreement
of the dip energy spectrum with observations (see Fig.~\ref{fig:dips}),
{\em (ii)} equality of all measured fluxes after the dip-based
energy recalibration (see Fig.~\ref{fig:recalib-all}) and {\em (iii)}
agreement of $\left\langle X_{\max}\right\rangle(E)$ with the bulk of
observations in the left panels of Fig.~\ref{fig:Xmax}.\\
At present the dip model is confirmed by the data of HiRes and TA
by strong proton-dominance at $E > 1$~EeV, and is in contradiction
with the Auger measurements of $\left\langle X_{\max}\right\rangle(E)$
and RMS at $E > 4$~EeV.
The key observation to accept or reject the dip model is the chemical
composition of UHECR at $E>1$ EeV. In the case of a substantial
admixture of nuclei in the spectrum ($>20\%$) the dip model should be
rejected.
\subsection{Mixed composition model}
\label{sec:mixedComposition}
The main concept of the mixed composition model (see Allard et al.\
\cite{mixed}) is based on the argument that any acceleration mechanism
operating in gas involves different species of nuclei in acceleration
process and thus the primary flux must have a mixed composition.

The basic features of the {\em mixed composition model} are as follows
\cite{mixed}: {\em (i)} In its basic versions \cite{mixed},
the source composition of extragalactic CRs is assumed to be almost the
same as for Galactic CRs, with protons and Helium being the dominant
components.  {\em (ii)} The source energy spectra are taken as
power-law with a generation index $\gamma_g=2.1 - 2.3$, with the
maximum acceleration energy assumed to be rigidity dependent, and with
cosmological evolution of the sources considered to be in
a wide range of regimes. {\em (iii)}  The sources usually are
assumed to be distributed homogeneously in the
universe.  The propagation is studied using Monte Carlo with
 nuclei photo-disintegration on the CMB and EBL. {\em (iv)}
At a first glance one may expect that a large number of free parameters,
such as generation index, parameters of cosmological evolution and
coefficients of source nuclei composition, can provide a broad variety
of observed mass compositions and spectra. However, as it was
demonstrated in \cite{mixed}, the predictions are very much constrained
due to photo-disintegration of nuclei on EBL and CMB radiations.
{\em (v)} The basic physics phenomena and their results are as follows.
Generically in the mixed models the mass composition becomes lighter at
$E > 10$~EeV, because intermediate and heavy nuclei are destroyed by the EBL
photons while protons survive. In principle this situation may change
only above $50$~EeV, when GZK cutoff in the proton spectrum appears,
while heavy nuclei, e.g.\ Iron, are still not photo-disintegrated by the CMB
photons and may dominate. In realistic cases the dominant component in
mixed models  are protons. {\em (vi)} Transition from Galactic to
extragalactic component in the mixed models
depends on the choice of parameters. In most models transition occurs
at the ankle, see Allard et al.\ in \cite{mixed}. However, in the
conceptually important paper by Allard, Olinto, Parizot (2007) from
\cite{mixed} it was emphasized that for strong source evolution and
flat generation spectra the intersection of Galactic and extragalactic
components occurs between $0.5$~EeV and $1$~EeV, i.e.\ at the second
knee, as in the dip model.

The dominance of protons  was
the reason why the $\left\langle X_{\max}\right\rangle(E)$ predicted by
mixed composition models is in a better agreement with HiRes data than
with PAO data. The observations of PAO show that mass composition becomes
heavier with increasing energy, and thus the existing calculations in
the framework  of mixed models agree better with the HiRes data.

However, with the recent proposal \cite{parizot} the power of the mixed
composition model for fitting the PAO data may change due to a possible
enhancement of the heavy nuclei production.
\section{Conclusions}
\label{sec:conclusions}
In this paper we limited our consideration by the models for {\em diffuse
fluxes} of UHECR. We will touch the problem of the produced neutrinos and  
detection of sources only shortly in the end of this section. The traditional 
interest for diffuse fluxes is given by the highest energies, namely by the
search for the GZK cutoff. Its unambiguous discovery, however, means
only that primaries are protons and exotic solutions are excluded.
The low-energy part of extragalactic CRs can give a key information
on the existence of pair-production dip, on propagation of CRs in
extragalactic magnetic fields and shed more light on the end of
Galactic CRs.

Therefore, the experimental studies in the transition region $(0.1 - 10)$
EeV are of paramount importance in this field of research, with the
mass composition measured by different methods being probably most
important task.

There are four working detectors which cover partially the above-mentioned
region: KASCADE-Grande \cite{KASCADE-G}, Tunka \cite{tunka},
Yakutsk \cite{Yak} and IceTop/IceCube \cite{IceTop}. There are also
projects to extend the observations of Telescope Array and PAO to low
energies, $(0.1 - 1)~$~EeV at TALE \cite{TALE} and at LE-Auger
\cite{LE-Auger}. The Auger detector has a great potential to
explore the low-energy region of the UHECR spectrum. At present there
are already two new detectors at PAO collecting data at this energy; 
High Elevation Auger Telescope (HEAT) \cite{HEAT}, detecting the fluorescent
light at higher elevation angles; and Auger Muons and Infill Ground Array 
(AMIGA) \cite{AMIGA}, for the detection of the EAS muon component.
These detectors, together with TALE, Tunka, Yakutsk and IceCube/IceTop
will provide information on all radiations from EAS, including
fluorescent and Cherenkov light, muons and radio radiation.
One may expect that in this way the present  controversy between
the mass compositions in the HiRes and Auger detectors will be
unambiguously solved. The recent measurement
\cite{Garcia-GamezICRC,Gazon2012} of the muon-production depth
$X_{\max}^{\mu}$ and maximum zenith angle $\theta_{\max}$ by
on-ground Auger detectors is an important step in this direction.

At the low-energy end of the UHECR the {\em energy spectra} are measured
with an unprecedented accuracy for cosmic ray physics. However, in fact
even in experiments with the same technique, like HiRes and Auger, the
{\em energy scales} are different due to systematic errors. However,
there is a physical 'standard candle' for the detector energy calibration,
given by the fixed energy position of the pair-production dip. The
recalibration factor $\lambda_{\rm cal}$ can be found by the spectrum
shift to the energy at which  agreement between observed and predicted
dips is the best (see \cite{BGGprd,Aletal} and  Bl\"umer at al.\ in 
\cite{GZK-rev}).

There are three models of transition: ankle, dip and mixed
composition one. The ankle model is excluded or severely
disfavoured by proton or Helium composition at energy $(1 -
3)$~EeV, i.e.\ below the ankle, where the particles have Galactic
origin. The mass composition at these energies will be reliably
measured by future low-energy detectors. However, the argument
against the ankle model obtained from Fig.~\ref{fig:Xmax} remains
to be valid independently from mass composition at  $(1 - 3)$~EeV,
unless it is very heavy.

A search for UHE neutrinos UHECR sources is outside the scope of this review.
However, a few remarks may be useful here.

In some particular dip models (see subsection \ref{sec:DipModel})
with proton-dominated  mass composition the cosmogenic neutrinos
can be detectable. The flux of cosmogenic neutrinos are severely
constrained by the electro-magnetic cascade upper limit
\cite{BeSm} and only in extreme cases \cite{BGKO,Ahlers} it can be
detectable, e.g.\ by JEM-EUSO \cite{jem-euso}. Another case of flux 
detectable by Ice-Cube at smaller energies $E \sim 10^{15}$~eV is 
given by cosmogenic neutrinos produced  at 'bright phase'\cite{BeBl}. 
UHE neutrinos can in principle indicate directions to the sources.
      
Even in the case when heavy nuclei dominate in the source
radiation, the protons are accelerated there too, and even small
produced flux can be detected from a nearby source. However, such
possibility not always exists. For example in the disappointing
model \cite{disapp} for interpretation of Auger results the
maximum energy of protons $E_{\max} \sim 4$~EeV is too small to
reach rectilinearly  detector from a source.
\begin{center}
Note added
\end{center}
Recently the KASCADE-Grande collaboration discovered (prd 87, 
081101, 2013) the light component at $10^{17} \lsim E \lsim 
10^{18}$~eV with a hard spectrum $\gamma = 2.79 \pm 0.08$. It 
consists of protons and Helium apparently of extragalactic 
origin. The transition from galactic, more steep, component 
occurs at $E \sim (1 - 2)\times 10^{17}$~eV and it becomes 
dominant at $E \gtrsim 10^{18}$~eV. The galactic origin of 
this component is disfavored by absence of observed anisotropy 
at $E \gtrsim 10^{18}$~eV. This observation favors the dip model 
and further disfavors ankle as  transition from galactic to 
extragalactic cosmic rays. 
\section{Acknowledgment}
This paper is based on the joint works and many discussions with
my co-authors Roberto Aloisio, Askhat Gazizov and Svetlana Grigorieva.
I am mostly grateful to them for efficient and pleasant collaboration.
I also learned much from many useful discussions with Pasquale Blasi,
Michael Kachelrie\ss, Johannes Knapp and Sergey Ostapchenko.
The work was partly supported by Ministry of Science and
Education of Russian Federation (agreement  8525).

\end{document}